\documentclass{article}

\usepackage[main, final, nonatbib]{neurips_2026}

\usepackage[utf8]{inputenc} 
\usepackage[T1]{fontenc}   
\usepackage{hyperref}      
\usepackage{url}            
\usepackage{booktabs}       
\usepackage{amsfonts}       
\usepackage{nicefrac}       
\usepackage{microtype}      
\usepackage{xcolor}         
\usepackage{fontawesome5}

\usepackage{multicol}
\usepackage{multirow}
\usepackage{graphicx}
\usepackage{colortbl}
\usepackage{amssymb}
\usepackage{xspace}
\usepackage{makecell}
\usepackage{bm}
\usepackage{subcaption}
\usepackage{enumitem}
\usepackage{amsmath}
\usepackage{wrapfig} 
\usepackage{subcaption}
\usepackage[table]{xcolor}
\usepackage{footmisc}

\newcommand{\mymodel}{FLASH}

\title{Rethinking Semantic ID Construction for Generative Recommendation: SimHash with Parallel Decoding and Semantic Alignment}

\author{
  Yuqing Liu$^{1,*}$\quad
  Huiyuan Chen$^{2,*,\dagger}$\quad
  Yibo Wang$^{1}$\quad
  Wooseong Yang$^{1}$\quad
  Philip S.\ Yu$^{1}$ \\[3pt]
  \normalfont
  $^{1}$University of Illinois Chicago\quad
  $^{2}$Amazon \\[2pt]
  \texttt{\{yliu363,ywang633,wyang73,psyu\}@uic.edu}\quad
  \texttt{huiyuach@amazon.com}
}

\begin{document}

\maketitle

{\renewcommand{\thefootnote}{*}\footnotetext{Equal contribution.\quad $^{\dagger}$Work done prior to joining Amazon.}}

\begin{abstract}
Semantic ID-based generative recommendation represents each item as a sequence of discrete tokens, enabling structured modeling of item semantics. A critical challenge is constructing semantic IDs that are both semantically expressive and computationally efficient. While recent approaches favor complex learned quantization, simple hashing-based methods such as SimHash are widely regarded as fundamentally inferior. In this work, we challenge this consensus by showing that the apparent performance gap does not stem from inherent limitations of hashing, but rather from a structural mismatch with autoregressive decoding, coupled with the inevitable information loss during rigid discretization. Based on this insight, we propose \mymodel, a two-stage framework that revitalizes training-free SimHash tokenization through parallel decoding and explicit semantic alignment. 
Despite its simplicity, \mymodel\ achieves state-of-the-art performance across multiple datasets without requiring any tokenizer training,
while exhibiting stronger generalization in cold-start scenarios. Notably, we demonstrate that semantic alignment acts as a universally effective mechanism across diverse paradigms. Our findings suggest that, with compatible decoding and semantic grounding, simple and efficient tokenizers can achieve performance comparable to complex learned counterparts in generative recommendation. 
Our project is available at \href{https://github.com/KevinC2015/Flash}{\textcolor{blue!60!black}{\faGithub\ \texttt{https://github.com/KevinC2015/Flash}}}.
\end{abstract}

\section{Introduction}
\label{sec:introduction}
Generative recommendation has emerged as a promising paradigm that frames next-item prediction as a sequence generation problem~\cite{li2024large,hstu,wang2024eager}, moving beyond traditional approaches that rely on learned atomic item embeddings and inner product-based retrieval~\cite{sasrec, bert4rec,rendle2012bpr,chen2024masked}.
A key direction within this paradigm is semantic ID (SID)-based recommendation, where each item is represented as a sequence of discrete tokens derived from its content, and a sequence model is trained to generate the semantic ID of the next item~\cite{hua2023index,zheng2024adapting,tiger, rpg, letter}. This shift to composable semantic codes enables knowledge sharing across items, improves generalization to rare or unseen items, and eliminates the need for large item-specific embedding tables. A central question is how to construct semantic IDs that are both expressive and well aligned with the decoding process.

Among existing approaches, residual quantization (RQ)~\cite{van2017neural,lee2022autoregressive} and product quantization (PQ)~\cite{jegou2010product} are two dominant strategies for semantic ID construction. 
RQ-based methods, represented by TIGER~\cite{tiger}, adopt a Residual-Quantized Variational Autoencoder (RQ-VAE)~\cite{lee2022autoregressive} to produce hierarchically dependent codes and rely on autoregressive (AR) decoding. However, they require training a dedicated quantizer and are typically limited to short semantic ID sequences (e.g., 4 tokens per item) due to the high inference cost of AR decoding, restricting their ability to capture fine-grained item semantics.
In contrast, PQ-based methods, represented by RPG~\cite{rpg}, apply optimized product quantization (OPQ)~\cite{ge2013optimized} to generate unordered codes and adopt parallel decoding, thereby improving efficiency and enabling longer semantic IDs.
Despite these differences, both approaches depend on learned quantizers, motivating the exploration of training-free alternatives such as Locality-Sensitive Hashing (LSH)~\cite{indyk1997locality,indyk1998approximate}, which maps item embeddings to discrete codes via random projections.
While training-free and efficient, this crude
discretization inevitably discards fine-grained semantic information compared to optimized quantizers. 
Furthermore, TIGER’s ablation study evaluates SimHash~\cite{charikar2002similarity} and finds that it performs significantly worse than RQ-VAE~\cite{tiger}. This observation has led to the prevailing view that LSH-based tokenization is less effective in generative recommendation.

\begin{wrapfigure}{r}{0.6\textwidth}  
    \centering
    \vspace{-12pt}  
    \includegraphics[width=0.58\textwidth]{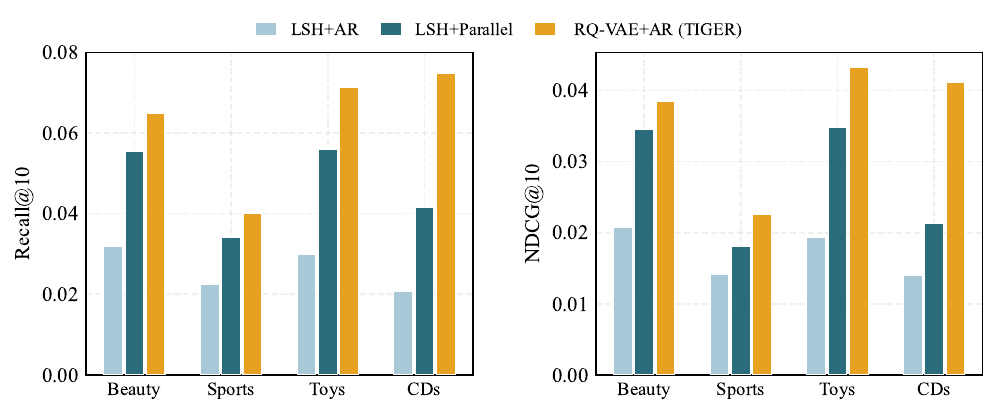}
    \vspace{-8pt}
    \caption{Preliminary study. Replacing autoregressive (AR) decoding with parallel decoding substantially improves LSH-based tokenization and narrows the gap with RQ-VAE. The remaining performance gap suggests that structural compatibility
alone is insufficient, motivating our introduction of semantic grounding.}
    \label{fig:prelim}
    \vspace{-7pt}  
\end{wrapfigure}

In this work, we revisit this conclusion and show that the performance gap can be attributed to a mismatch between tokenization structure and decoding paradigm, rather than the tokenization itself.
Specifically, SimHash produces codes that do not admit a natural ordering and lack inherent dependencies across positions, whereas autoregressive decoding imposes sequential dependencies.  Our preliminary study (Figure~\ref{fig:prelim}; see Appendix~\ref{sec:app_implement} for detailed setup) shows that replacing autoregressive decoding with parallel decoding significantly improves the performance of LSH-based tokenization, indicating that such incompatibility introduces harmful inductive bias.
However, this improvement alone does not fully close the gap with learned quantization methods, suggesting that additional factors beyond decoding compatibility play an important role.

Building on this insight, we derive two design principles for effective semantic ID-based generative recommendation: 
(1) \textbf{structural compatibility}, which requires the decoding paradigm to align with the dependency structure induced by the tokenization method. 
Crucially, enforcing this compatibility via parallel decoding inherently removes the inference bottleneck and ID length constraint of autoregressive generation, enabling the use of substantially longer semantic ID sequences to boost representational capacity.
(2) \textbf{semantic grounding}, which explicitly compensates for the loss of fine-grained information introduced by discrete tokenization, particularly for training-free methods. 

Guided by these principles, we propose \textbf{FLASH} (\textbf{\underline{F}}ast \textbf{\underline{L}}LM-\textbf{\underline{A}}ligned \textbf{\underline{S}}im\textbf{\underline{H}}ash for Generative Recommendation), a simple yet effective two-stage framework that integrates training-free SimHash tokenization with parallel decoding and LLM-based semantic alignment. 
In the first stage, items are mapped to semantic IDs using SimHash over pre-computed LLM text embeddings, requiring no tokenizer training. 
In the second stage, a sequence model encodes user interaction histories. 
To ensure structural compatibility with SimHash codes, it predicts the semantic ID of the target item using parallel decoding under a multi-token prediction (MTP) objective. 
To compensate for the semantic information loss introduced by hashing, we incorporate an alignment loss that anchors the learned representations to the original LLM embedding space during training.

Overall, our results demonstrate that effective semantic ID-based generative recommendation depends on both structural compatibility and semantic grounding, and that LSH-based methods can perform strongly with appropriate decoding and semantic supervision.
Beyond this core insight, \mymodel\ combines competitive recommendation performance with high efficiency through training-free tokenization, parallel inference, and support for long semantic IDs.
Furthermore, \mymodel\ exhibits strong generalization in cold-start settings.
Our contributions are summarized as follows:
\vspace{-8pt}
\begin{itemize}
    \setlength{\itemsep}{-2pt}    
    \item By revisiting prior conclusions about the limited effectiveness of LSH-based tokenization, we derive two key design principles for semantic ID-based generative recommendation: structural compatibility and semantic grounding.
    \item We propose \mymodel, a simple yet effective unified framework that instantiates these principles by integrating training-free SimHash tokenization, parallel decoding, and LLM-based semantic alignment.
    \item We demonstrate that FLASH achieves competitive performance across multiple datasets without requiring tokenizer training, and exhibits superior generalization capabilities in cold-start scenarios. Furthermore, explicit semantic alignment serves as a general and effective regularization strategy across different quantization and decoding paradigms.
\end{itemize}

\section{Methodology}
\label{sec:methodology}

\begin{figure}[t]
    \centering
\includegraphics[width=\linewidth]{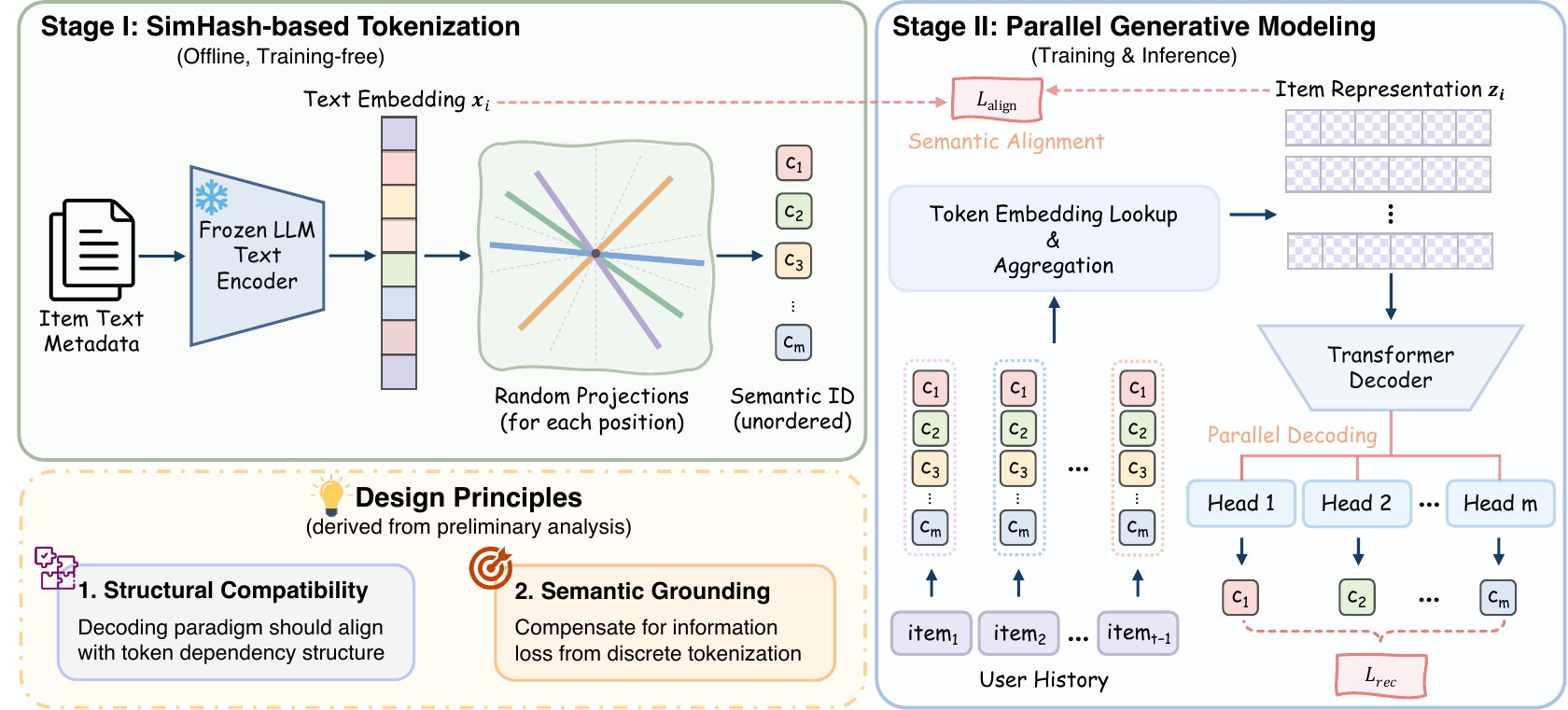}
    \caption{Overview of \mymodel. Stage~I constructs semantic IDs using SimHash over pre-computed LLM text embeddings in a training-free manner.
    Stage~II performs parallel generative modeling with a multi-token prediction objective and incorporates an alignment loss for semantic grounding.
The design is guided by structural compatibility and semantic grounding, aligning the decoding paradigm with the tokenization structure while compensating for information loss.}
    \label{fig:framework}
\end{figure}

We formulate the task as a sequential recommendation problem. Given a user's historical interaction sequence $\mathcal{S} = (i_1, \ldots, i_{t-1})$ over an item set $\mathcal{I}$, the goal is to predict the next item $i_t$.
In semantic ID-based generative recommendation, each item $i \in \mathcal{I}$ is represented by a tuple of discrete tokens $c^{(i)} = (c_1^{(i)}, \ldots, c_m^{(i)})$, where $m$ is the number of codebooks, and each token $c_j^{(i)}\in \mathcal{C}_{j}=\{1,\ldots, M\}$ is drawn from the $j$-th codebook. The recommendation task is thus reformulated as generating the semantic ID $(c_1^{(t)}, \ldots, c_m^{(t)})$ of the next item.
Guided by the design principles of structural compatibility and semantic grounding, we propose \mymodel, a two-stage framework consisting of: (1) training-free SimHash tokenization (\S\ref{sec:stage1}), and (2) parallel generative modeling (\S\ref{sec:stage2}). In particular, Stage II enforces structural compatibility through parallel decoding (\S\ref{sec:parallel}) and achieves semantic grounding via an alignment objective (\S\ref{sec:alignment}). An overview of the framework is shown in Figure~\ref{fig:framework}.

\subsection{Stage I: SimHash-based Semantic ID Construction}
\label{sec:stage1}

\textbf{Semantic ID Construction.}
We construct semantic IDs by applying SimHash~\cite{charikar2002similarity} to pre-computed LLM text embeddings. 
Each item $i$ is first encoded into a dense representation $\mathbf{x}_i \in \mathbb{R}^{d_{\text{text}}}$ using a pre-trained LLM based on its textual metadata (e.g., title, description, category).
These embeddings are computed once offline and kept fixed.
For each code position $j \in \{1, \ldots, m\}$, we sample $B = \lceil \log_2 M \rceil$ random hyperplanes $\{\mathbf{r}_{j,b}\}_{b=1}^B \subset \mathbb{R}^{d_{\text{text}}}$ from a standard Gaussian distribution, where $M$ is the codebook size.
The code at position $j$ for item $i$ is obtained by projecting $\mathbf{x}_i$ onto these independent sets of hyperplanes and converting the resulting sign bits into an integer:
\begin{equation}
    c_j^{(i)} = \left(\sum_{b=1}^{B} \mathbf{1}\!\left[\mathbf{x}_i \cdot \mathbf{r}_{j,b} > 0\right] \cdot 2^{b-1}\right) \bmod M + 1.
    \label{eq:simhash}
\end{equation}
We restrict the codebook size $M$ to be a power of 2, so that the modulo mapping distributes binary sequences more uniformly across the codebook. 
The modulo operation yields values in $\{0, \ldots, M-1\}$, and we shift the result by $+1$ to obtain 1-indexed semantic IDs. The hyperplanes are fixed and shared across all items, resulting in a training-free tokenization process.
 
\textbf{Structural Properties and Design Implications.}
SimHash-based semantic IDs exhibit structural properties that inform model design. 
First, codes across positions are generated individually without residual dependencies.
Moreover, SimHash imposes no intrinsic ordering over positions, making autoregressive decoding structurally mismatched. 
Finally, the discretization process introduces information loss despite preserving coarse-grained similarity. 
These properties motivate the design of \mymodel\ in \S\ref{sec:stage2}, where parallel decoding aligns with unordered structure and semantic alignment compensates for information loss.

\subsection{Stage II: Parallel Generative Modeling}
\label{sec:stage2}
    
\subsubsection{Semantic ID-based Item Representation and Sequence Modeling}
\label{sec:input}
\textbf{Item Representation.} 
Given an item with semantic ID $(c_1, \ldots, c_m)$, we associate each position $j$ with a learnable codebook embedding table 
$\mathbf{E}^{(j)} \in \mathbb{R}^{M \times d}$, where $d$ is the hidden embedding dimension.
Each code $c_j$ is mapped to its corresponding embedding 
$\mathbf{e}^{(j)}_{c_j} \in \mathbb{R}^d$ via table lookup.
The resulting code embeddings $\{\mathbf{e}^{(j)}_{c_j}\}_{j=1}^m$ are aggregated into a single item representation:
\begin{equation}
    \mathbf{z} = \mathrm{Aggr}(\mathbf{e}^{(1)}_{c_1}, \ldots, \mathbf{e}^{(m)}_{c_m}),
    \label{eq:item_representation}
\end{equation}
where $\mathrm{Aggr}(\cdot)$ denotes an aggregation function.
In this work, we instantiate $\mathrm{Aggr}(\cdot)$ as a linear projection over the concatenation of code embeddings, mapping the $md$-dimensional input to a $d$-dimensional representation.
This design treats all code positions uniformly at the structural level, consistent with the unordered nature of SimHash-based SIDs. 
At the same time, it allows the model to learn distinct contributions from different codebook embeddings.
Furthermore, to mitigate over-reliance on specific code positions and improve robustness, we apply position-level codebook dropout during training by randomly masking entire code positions (i.e., removing the corresponding code embeddings) and rescaling the remaining embeddings to preserve their expected magnitude.

\textbf{Sequence Modeling.}
The sequence of aggregated item representations $\{\mathbf{z}_1, \ldots,\mathbf{z}_{t-1}\}$ (derived from Eq.~\ref{eq:item_representation}) is encoded by a Transformer backbone, producing contextual hidden states $\{\mathbf{h}_i\}_{i=1}^{t-1}$, where each $\mathbf{h}_i \in \mathbb{R}^d$ summarizes the user interaction history up to position $i$. We use the final hidden state $\mathbf{h}_{t-1}$ as the representation of the user interaction history for next-item prediction.

\subsubsection{Parallel Decoding for Structural Compatibility}
\label{sec:parallel}
Given an encoded user interaction history $\mathbf{h}_{t-1}$, the recommendation task is to model the conditional probability distribution over the semantic ID of the next item: $P((c_1^{(t)}, \ldots, c_m^{(t)}) \mid \mathbf{h}_{t-1})$.

A key design principle in generative recommendation is the alignment between the decoding paradigm and the dependency structure of semantic IDs. 
SimHash-based semantic IDs do not impose sequential dependencies or an intrinsic ordering across code positions.
Motivated by this unordered structure, we adopt a factorized formulation that enables all $m$ code positions to be predicted simultaneously:
\begin{equation}
    P((c_1^{(t)}, \ldots, c_m^{(t)}) \mid \mathbf{h}_{t-1})
= \prod_{j=1}^{m} p_{j}(c_j^{(t)} \mid \mathbf{h}_{t-1}).
\end{equation}
We parameterize each factor using a dedicated prediction head for code position $j$, yielding $m$ parallel heads. 
This corresponds to a multi-token prediction (MTP) setting, where all code positions are predicted in parallel.
Concretely, each prediction head produces a probability distribution over $M$ codes at position $j$ via dot-product similarity with a learnable temperature $\tau_j$:
\begin{equation}
    p_{j}(c \mid \mathbf{h}_{t-1}) =
        \frac{
            \exp\!\left({f}_j(\mathbf{h}_{t-1})^\top \cdot \mathbf{e}^{(j)}_{c}\,/\,\tau_j\right)
        }{
            \sum_{c'\in \mathcal{C}_{j}}
            \exp\!\left({f}_j(\mathbf{h}_{t-1})^\top \cdot \mathbf{e}^{(j)}_{c'}\,/\,\tau_j\right)
        },
\end{equation}
where each $f_j(\cdot)$ is a lightweight MLP projection head~\cite{chen2020simple, cai2024medusa} that maps contextual representations to the codebook space $\mathbf{E}^{(j)}$, shared between input and output.
Here $c \in \mathcal{C}_j$ denotes a candidate code, and the target probability is obtained by evaluating at $c = c_j^{(t)}$.
Given a target item with SID $(c_1^{(t)}, \ldots, c_m^{(t)})$, the recommendation loss is defined as the average cross-entropy across all $m$ heads:
\begin{equation}
    \mathcal{L}_{\text{rec}} = -\frac{1}{m} \sum_{j=1}^{m} \log p_{j}(c_j^{(t)} \mid \mathbf{h}_{t-1}).
    \label{eq:rec_loss}
\end{equation}
This MTP objective jointly optimizes all code positions under the factorized formulation.
Moreover, parallel decoding supports longer semantic ID sequences by avoiding autoregressive constraints.

\subsubsection{Semantic Alignment for Information Loss}
\label{sec:alignment}
While SimHash preserves coarse-grained similarity through its locality-sensitive property, the discretization process inevitably discards fine-grained semantic information.
Unlike RQ-VAE, which implicitly anchors item representations through its reconstruction objective, SimHash lacks the semantic grounding signal.
To compensate for this loss, we introduce a semantic alignment objective that regularizes the input representation space toward the original LLM embedding space, thereby preserving semantic consistency. 
Specifically, we introduce a trainable linear projector $\phi\colon \mathbb{R}^{d_{\text{text}}} \to \mathbb{R}^d$ and align each input item representation in the sequence with its corresponding (frozen) LLM text embedding by minimizing the negative cosine similarity, providing an explicit semantic anchor:
\begin{equation}
    \mathcal{L}_{\text{align}} =
        -\frac{1}{|\mathcal{S}|}
        \sum_{i \in \mathcal{S}}
        \cos\!\left(\mathbf{z}_{i},\;\phi(\mathbf{x}_{i})\right).
    \label{eq:align_loss}
\end{equation}
The overall training objective can be represented as
\begin{equation}
    \mathcal{L} =
        \mathcal{L}_{\text{rec}}
        + \lambda_{\text{align}}\,\mathcal{L}_{\text{align}},
    \label{eq:total_loss}
\end{equation}
where $\lambda_{\text{align}}$ is the hyperparameter.
The alignment loss improves semantic fidelity, while the recommendation loss supervises the predictive distribution over next-item SIDs, providing complementary signals for representation quality and predictive accuracy. Overall, the design enforces structural compatibility at the architectural level and semantic grounding at the representation level.

\subsection{Efficient Inference}
\label{sec:inference}
 
At inference, given the final hidden state $\mathbf{h}_{t-1}$, we rank all candidate items without enumerating all possible SID combinations, and return the top-$k$ items by score as predictions.

Parallel decoding enables item scoring by aggregating position-wise probabilities, with per-head probabilities computed once and cached.
Specifically, we first compute the probability distribution over all $M$ codebook entries at each position, following the cached logit strategy in~\cite{rpg}:
\begin{equation}
    \mathbf{q}^{(j)} =
        \operatorname{softmax}\!\left(
            \mathbf{E}^{(j)} \cdot f_j({\mathbf{h}_{t-1}}) \,/\,\tau_j
        \right) \in \mathbb{R}^M.
    \label{eq:logit_cache}
\end{equation}
This decouples the computation from the number of items $N$, as the per-head probabilities are computed once and reused for all items, resulting in $O(mMd)$ time complexity.
Given these cached probabilities, the score of a candidate item can be computed by summing the corresponding log-probabilities across code positions. 
Specifically, for an item $i$ with semantic ID $(c_1, \dots, c_m)$, its score is given by
\begin{equation}
    \operatorname{score}(i) = \sum_{j=1}^m \log \mathbf{q}^{(j)}_{c_j}.
    \label{eq:item_score}
\end{equation}
Consequently, inference reduces to an $O(Nm)$ gather-and-sum operation, leading to a total cost of $O(mMd + Nm)$, compared to $O(Nmd)$ for naive per-item similarity evaluation.

\section{Experiments}
\label{sec:experiments}
\subsection{Experimental Setup}
\label{sec:experiment_setup}
\textbf{Datasets.}
We evaluate \mymodel\ on four widely used categories from the Amazon Product Reviews dataset~\cite{dataset}: “Beauty”, “Toys and Games” (Toys), “Sports and Outdoors” (Sports), and “CDs and Vinyl” (CDs). We follow standard preprocessing and evaluation protocols commonly used in prior work~\cite{tiger,rpg}. Detailed dataset statistics and preprocessing steps are provided in Appendix~\ref{sec:app_dataset}. 

\textbf{Baselines.}
We compare \mymodel\ against three categories of baselines:
(1)~\textit{Item ID-based methods}: Caser~\cite{caser}, GRU4Rec~\cite{gru4rec}, HGN~\cite{hgn}, BERT4Rec~\cite{bert4rec}, SASRec~\cite{sasrec}, and S$^3$-Rec~\cite{s3rec}, which represent items using unique identifiers and learn embeddings from interaction data;
(2)~\textit{RQ-based semantic ID}: TIGER~\cite{tiger}, LETTER~\cite{letter}, and DIGER~\cite{diger}, which construct semantic IDs via learned residual quantization and model them with autoregressive decoding;
and (3)~\textit{PQ-based semantic ID}: RecJPQ~\cite{recjpq}, VQ-Rec~\cite{vqrec}, and RPG~\cite{rpg}, which derive semantic IDs using product quantization and differ in modeling strategies.
See Appendix~\ref{sec:app_baselines} for detailed descriptions of baselines. 

\textbf{Evaluation.}
We evaluate recommendation performance using two widely adopted metrics, Recall (R@$K$) and NDCG (N@$K$). Following prior work~\cite{tiger}, we set~$K=5,10$ by default.

\textbf{Implementation Details.}
We adopt most baseline results from TIGER~\cite{tiger} and RPG~\cite{rpg}, and implement the remaining methods using their publicly available source code. For all methods, hyperparameters are carefully tuned following the original papers for fair comparison. 
For \mymodel, we perform a grid search over the number of codebooks $m \in \{4, 8, 16, 32, 64\}$, codebook size $M \in \{128, 256, 512, 1024\}$, and alignment loss weight $\lambda_{\text{align}} \in [0, 1]$. The learning rate is tuned within $[0.001, 0.01]$, with a fixed position-level codebook dropout ratio of $0.2$. Models are trained with a batch size of $256$ for up to $300$ epochs, 
with early stopping if validation performance does not improve for $20$ consecutive epochs.
Additional details are provided in Appendix~\ref{sec:app_implement}.

\subsection{Overall Performance}
We compare \mymodel\ with three categories of baselines across four public datasets and report the overall performance in Table~\ref{tab:overall}. 
\mymodel\ consistently achieves the best performance on all datasets and metrics, with relative improvements on N@5 ranging from $5.51\%$ to $14.46\%$, demonstrating its effectiveness and robustness.

\begin{figure}[tb]
    \centering
    \begin{subfigure}{0.47\textwidth}
        \centering
        \includegraphics[width=\linewidth]{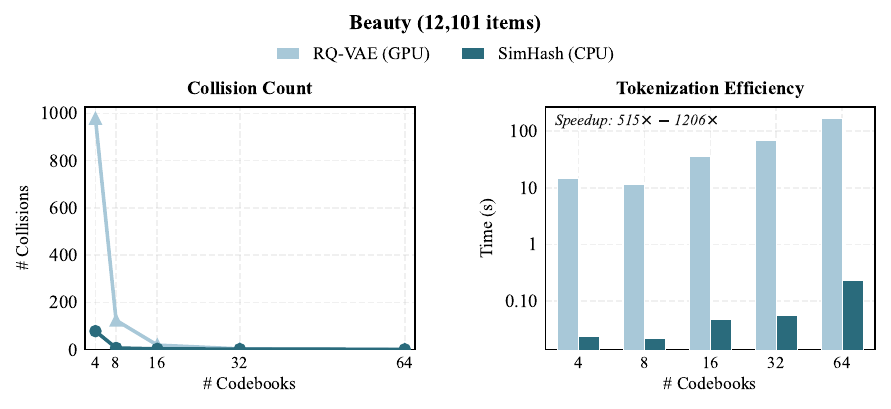}
    \end{subfigure}
    \hfill
    \begin{subfigure}{0.47\textwidth}
        \centering
        \includegraphics[width=\linewidth]{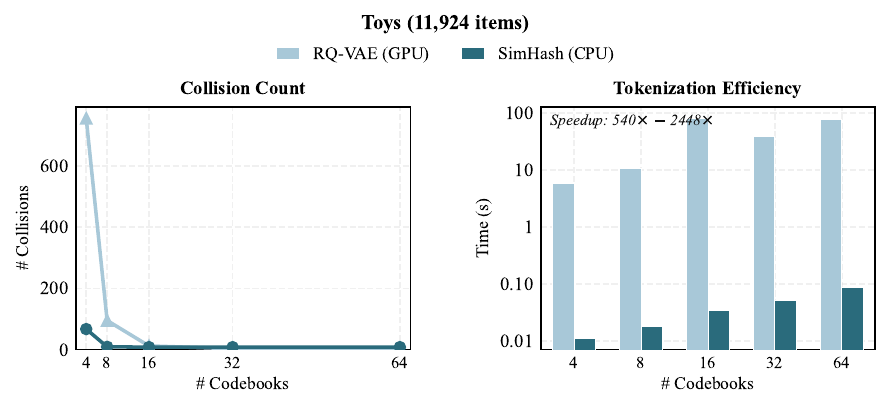}
    \end{subfigure}
    \vspace{0.5em}  
    \begin{subfigure}{0.47\textwidth}
        \centering
        \includegraphics[width=\linewidth]{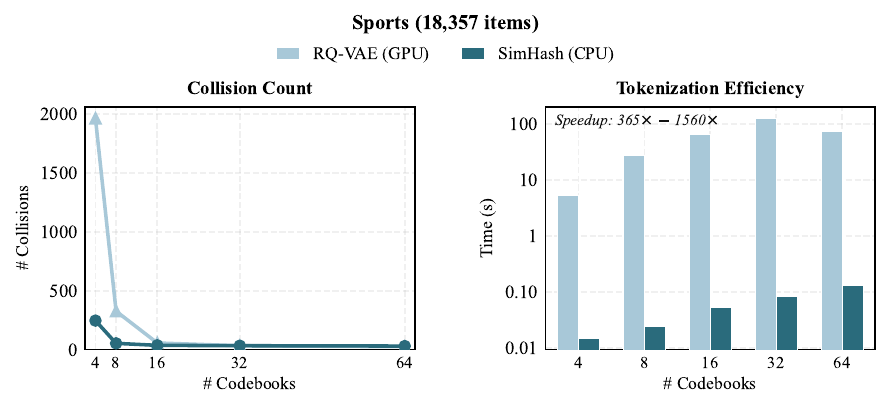}
    \end{subfigure}
    \hfill
    \begin{subfigure}{0.47\textwidth}\centering\includegraphics[width=\linewidth]{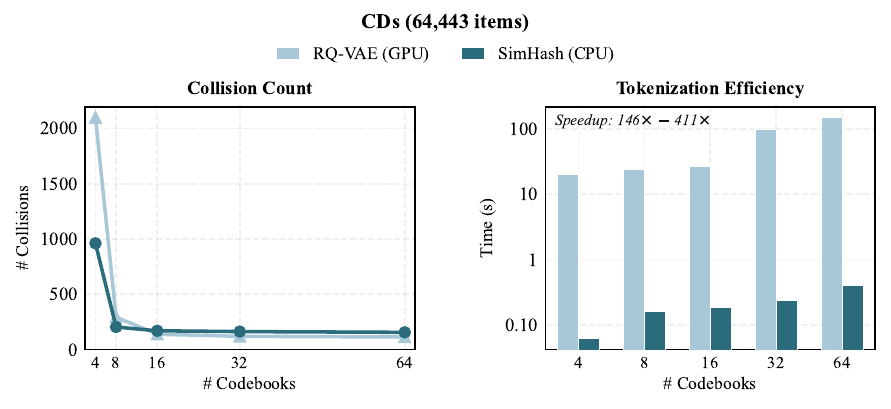}
    \end{subfigure}
    \vspace{-10pt}
    \caption{Collision statistics and tokenization time for SimHash and RQ-VAE across different numbers of codebooks with fixed codebook size of 256.}
    \label{fig:collision}
    \vspace{-12pt}
\end{figure}

Compared to item ID-based methods, \mymodel\ achieves substantial improvements across all datasets, suggesting that representations grounded in item semantic information provide more informative signals than discrete item identifiers for recommendation.
Within the semantic ID-based paradigm, \mymodel\ outperforms both RQ-based and PQ-based methods.
For example, compared to RPG, the strongest PQ-based baseline, \mymodel\ achieves gains of $5.18\%$ on R@10 and $11.02\%$ on N@10 on the Toys dataset.
This result is particularly notable given that \mymodel\ employs a training-free tokenization scheme, in contrast to learned quantization approaches that require additional training objectives and model parameters. 
\mymodel\ is also more efficient than TIGER due to its training-free tokenization and parallel prediction (see Figure~\ref{fig:collision} and Appendix~\ref{sec:app_efficiency}).

The superior performance of \mymodel\ can be attributed to three key design choices. First, SimHash enables efficient semantic ID construction while preserving coarse-grained semantic similarity from LLM-based embeddings. Second, parallel decoding aligns naturally with the unordered structure of LSH-based tokens, avoiding the mismatch introduced by autoregressive decoding. Third, the semantic alignment objective compensates for information loss introduced by discrete hashing, retaining fine-grained semantic signals.
Together, these findings highlight the importance of structural compatibility and semantic grounding in semantic ID-based generative recommendation.

\begin{table*}[t!]
  \centering
  \caption{Overall performance comparison among baselines and \mymodel\ across four datasets. The best result is in \textbf{bold} and the second-best result is \underline{underlined}. \mymodel\ achieves statistically significant improvements over the best baseline under a paired $t$-test ($p<0.05$).
  }
  \label{tab:overall}
  \setlength{\tabcolsep}{1mm}{
  \resizebox{\textwidth}{!}{
  \begin{tabular}{@{}p{1.6cm}@{\hspace{1mm}}cccccccccccccccc@{}}
    \toprule
    \multicolumn{1}{p{1.6cm}}{\multirow{2.5}{*}{\textbf{Model}}} & \multicolumn{4}{c}{\textbf{Beauty}} & \multicolumn{4}{c}{\textbf{Toys and Games}} & \multicolumn{4}{c}{\textbf{Sports and Outdoors}} & \multicolumn{4}{c}{\textbf{CDs and Vinyl}} \\
    \cmidrule(l){2-5} \cmidrule(l){6-9} \cmidrule(l){10-13} \cmidrule(l){14-17}
    & R@5 & N@5 & R@10 & N@10 & R@5 & N@5 & R@10 & N@10 & R@5 & N@5 & R@10 & N@10 & R@5 & N@5 & R@10 & N@10 \\
    \midrule\midrule
    \multicolumn{17}{@{}c}{\textit{\quad Item ID-based}} \\
    \midrule
    Caser    & 0.0205 & 0.0131 & 0.0347 & 0.0176 & 0.0166 & 0.0107 & 0.0270 & 0.0141 & 0.0116 & 0.0072 & 0.0194 & 0.0097 & 0.0116 & 0.0073 & 0.0205 & 0.0101 \\
    GRU4Rec  & 0.0164 & 0.0099 & 0.0283 & 0.0137 & 0.0097 & 0.0059 & 0.0176 & 0.0084 & 0.0129 & 0.0086 & 0.0204 & 0.0110 & 0.0195 & 0.0120 & 0.0353 & 0.0171\\
    HGN      & 0.0325 & 0.0206 & 0.0512 & 0.0266 & 0.0321 & 0.0221 & 0.0497 & 0.0277 & 0.0189 & 0.0120 & 0.0313 & 0.0159 & 0.0259 & 0.0153 & 0.0467 & 0.0220\\
    BERT4Rec & 0.0203 & 0.0124 & 0.0347 & 0.0170 & 0.0116 & 0.0071 & 0.0203 & 0.0099 & 0.0115 & 0.0075 & 0.0191 & 0.0099 & 0.0326 & 0.0201 & 0.0547 & 0.0271\\
    SASRec   & 0.0387 & 0.0249 & 0.0605 & 0.0318 & 0.0463 & 0.0306 & 0.0675 & 0.0374 & 0.0233 & 0.0154 & 0.0350 & 0.0192 & 0.0351 & 0.0177 & 0.0619 & 0.0263\\
    S$^3$-Rec & 0.0387 & 0.0244 & 0.0647 & 0.0327 & 0.0443 & 0.0294 & 0.0700 & 0.0376 & 0.0251 & 0.0161 & 0.0385 & 0.0204 & 0.0213 & 0.0130 & 0.0375 & 0.0182 \\
    \midrule
    \multicolumn{17}{@{}c}{\textit{\quad RQ-based Semantic ID}} \\
    \midrule
    TIGER  & 0.0454 & 0.0321 & 0.0648 & 0.0384 & 0.0521 & 0.0371 & 0.0712 & 0.0432 & 0.0264 & 0.0181 & 0.0400 & 0.0225 & 0.0492 & 0.0329 & \underline{0.0748} & 0.0411 \\
    LETTER & 0.0489 & 0.0364 & 0.0689 & 0.0402 & 0.0544 & 0.0379 & 0.0727 & 0.0437 & 0.0293 & 0.0187 & 0.0420 & 0.0232 & 0.0481 & 0.0314 & 0.0723 & 0.0398 \\
    DIGER & 0.0503 & 0.0370 & 0.0692 & 0.0421 & 0.0540 & 0.0383 & 0.0719 & 0.0442 & 0.0284 & 0.0192 & 0.0388 & 0.0241 & 0.0470 & 0.0312 & 0.0717 & 0.0369 \\
    \midrule
    \multicolumn{17}{@{}c}{\textit{\quad PQ-based Semantic ID}} \\
    \midrule
    RecJPQ & 0.0311 & 0.0167 & 0.0482 & 0.0222 & 0.0331 & 0.0182 & 0.0484 & 0.0231 & 0.0141 & 0.0076 & 0.0220 & 0.0102 & 0.0075 & 0.0046 & 0.0138 & 0.0066 \\
    VQ-Rec & 0.0457 & 0.0317 & 0.0664 & 0.0383 & 0.0497 & 0.0346 & 0.0737 & 0.0423 & 0.0208 & 0.0144 & 0.0300 & 0.0173 & 0.0352 &  0.0238 & 0.0520 & 0.0292 \\
    RPG    & \underline{0.0550} & \underline{0.0381} & \underline{0.0809} & \underline{0.0464} & \underline{0.0592} & \underline{0.0401} & \underline{0.0869} & \underline{0.0490} & \underline{0.0314} & \underline{0.0216} & \underline{0.0463} & \underline{0.0263} & \underline{0.0498} & \underline{0.0338} & 0.0735 & \underline{0.0415} \\
    \midrule
    \multicolumn{17}{@{}c}{\textit{\quad LSH-based Semantic ID}} \\
    \midrule
    \textbf{\mymodel} & \textbf{0.0576}	 & \textbf{0.0402}	 & \textbf{0.0832}
 & \textbf{0.0489}	 & \textbf{0.0650}	 & \textbf{0.0459}	 & \textbf{0.0914}
 & \textbf{0.0544}	 & \textbf{0.0337}	 & \textbf{0.0237}	 & \textbf{0.0487}
 & \textbf{0.0285}	 & \textbf{0.0536}	 & \textbf{0.0370}	 & \textbf{0.0769}      & \textbf{0.0446} \\
    \midrule
    \rowcolor{gray!12}
    \textit{Improv.} & \textit{+4.73\%} & \textit{+5.51\%} & \textit{+2.84\%} & \textit{+5.39\%} & \textit{+9.80\%} & \textit{+14.46\%} & \textit{+5.18\%} & \textit{+11.02\%} & \textit{+7.32\%} & \textit{+9.72\%} & \textit{+5.18\%} & \textit{+8.37\%} & \textit{+7.63\%} & \textit{+9.47\%} & \textit{+2.81\%} & \textit{+7.47\%} \\
    \bottomrule
    \bottomrule
  \end{tabular}
  }}
\end{table*}

\begin{table}[t!]
\small
\centering
\caption{Effect of tokenization methods under unified parallel decoding and semantic alignment. $T$ denotes wall-clock tokenization time (s). $^{\dagger}$RQ-VAE and RQ-KMeans are measured on GPU; others on CPU. Best recommendation results are in \textbf{bold} and second-best are \underline{underlined}.}
\vspace{2pt}
\label{tab:ablation_tokenization}
\resizebox{\textwidth}{!}{
\begin{tabular}{lcccccccccccc}
\toprule
& \multicolumn{3}{c}{Beauty}
& \multicolumn{3}{c}{Toys}
& \multicolumn{3}{c}{Sports}
& \multicolumn{3}{c}{CDs} \\
\cmidrule(lr){2-4}\cmidrule(lr){5-7}\cmidrule(lr){8-10}\cmidrule(lr){11-13}
Tokenizer & R@10 & N@10 & T (s) $\downarrow$ & R@10 & N@10 & T (s) $\downarrow$
& R@10 & N@10 & T (s) $\downarrow$ & R@10 & N@10 & T (s) $\downarrow$ \\
\midrule
Random & 0.0432   & 0.0261 & 0.0177
& 0.0612          & 0.0324 & 0.0144
& 0.0361          & 0.0198 & 0.0220
& 0.0197          & 0.0098 & 0.0353 \\
RQ-VAE$^{\dagger}$& \underline{0.0832}  & 0.0483 & 67.08
& 0.0856          & 0.0481 & 77.41
& 0.0466 & 0.0278 & 73.65
& 0.0751          & 0.0438 &  147.26 \\  
RQ-KMeans$^{\dagger}$ &  0.0812  &  0.0462 & 8.19
&     0.0861    &  0.0490 & 12.07
&  0.0457 &  0.0263 & 20.90
&  0.0760   & 0.0441 & 132.87 \\ 
OPQ& \textbf{0.0851}          & \textbf{0.0498} & 292.89
& \underline{0.0905}          & \underline{0.0523} & 111.08
& 0.0460          & 0.0265 & 151.90
& \textbf{0.0782} & \textbf{0.0455} & 162.98 \\
SimHash& \underline{0.0832} & \underline{0.0489} & 0.0556
& \textbf{0.0914} & \textbf{0.0544} & 0.0869
& \textbf{0.0487} & \textbf{0.0285} & 0.1340
& \underline{0.0769} & \underline{0.0446} & 0.4608 \\
\bottomrule
\end{tabular}}
\end{table}
\vspace{-10pt}

\subsection{Ablation Study}
\subsubsection{Analysis of Structural Compatibility}
\label{sec:ablation_tokenizer}

To investigate the impact of structural compatibility, we replace SimHash with alternative tokenization methods while keeping the downstream sequence modeling pipeline fixed, following the dataset-specific configuration of \mymodel\ reported in Table~\ref{tab:best_hyperparameters}.
Specifically, we consider \textbf{Random} code assignment with no semantic basis and three learned quantizers: \textbf{RQ-VAE}, \textbf{RQ-KMeans}, and \textbf{OPQ}.
The results are reported in Table~\ref{tab:ablation_tokenization}.

SimHash achieves competitive or superior performance compared to learned quantization methods across most datasets, despite requiring no tokenizer training.
Random assignment performs substantially worse, with over $40\%$ lower N@10 than SimHash on average, showing that structural compatibility alone is insufficient without meaningful semantic information. 
Among learned tokenizers, OPQ generally outperforms the residual methods under the same downstream modeling pipeline, further suggesting the importance of compatibility between the tokenization structure and the decoding paradigm. 
However, the relatively small performance gaps among learned tokenizers and SimHash indicate that parallel decoding and semantic grounding can partially mitigate differences in tokenization.
Notably, SimHash substantially reduces tokenization cost, requiring less than $0.5$ seconds on CPU across all datasets, compared with tens to hundreds of seconds for learned quantizers.

To further characterize the tokenizers beyond recommendation performance, we compare the code collision behavior of SimHash and RQ-VAE across multiple datasets and codebook configurations. As shown in Figure~\ref{fig:collision}, the number of collisions for SimHash is consistently low across all datasets and decreases as the number of codebooks increases. For example, on the Beauty dataset, collisions reduce from 78 at $m=4$ to only 1 at $m=64$. 
A similar trend is observed across larger datasets such as CDs, where collisions remain a small fraction of the total number of items even at scale.
Importantly, the collision count for SimHash is comparable to, and in some cases lower than, that of RQ-VAE under the same configurations.
Combined with its substantially lower tokenization cost, these results show that SimHash retains the advantages of simplicity and efficiency without sacrificing code distinguishability.

\subsubsection{Analysis of Semantic Grounding}
\label{sec:ablation_align}

To evaluate the effect of semantic grounding, we compare model performance with and without the alignment objective. For clarity, we present results on the Beauty and Toys datasets in the main text, while results on the remaining datasets are provided in Appendix~\ref{sec:app_alignment} and exhibit similar trends.

\begin{figure}[t!]
    \centering
    \begin{subfigure}{0.24\textwidth}
        \centering
        \includegraphics[width=\linewidth]{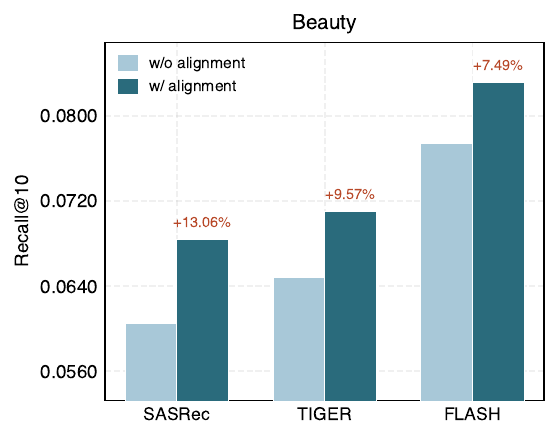}
    \end{subfigure}
    \hfill
    \begin{subfigure}{0.24\textwidth}
        \centering
        \includegraphics[width=\linewidth]{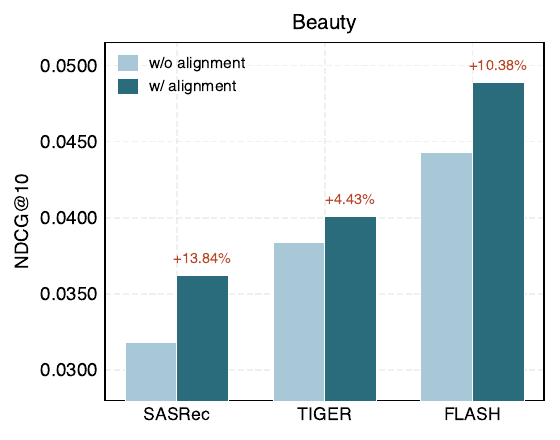}
    \end{subfigure}
    \hfill
    \begin{subfigure}{0.24\textwidth}
        \centering
        \includegraphics[width=\linewidth]{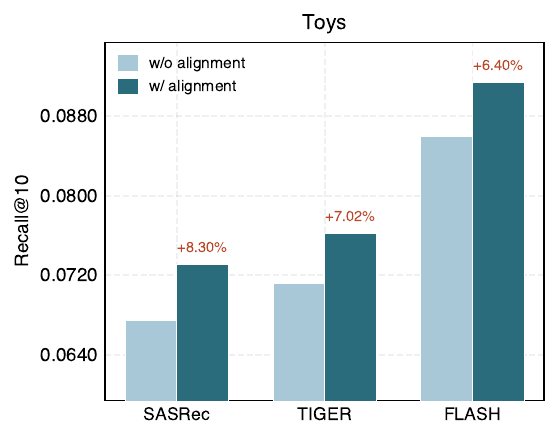}
    \end{subfigure}
    \hfill
    \begin{subfigure}{0.24\textwidth}
        \centering
        \includegraphics[width=\linewidth]{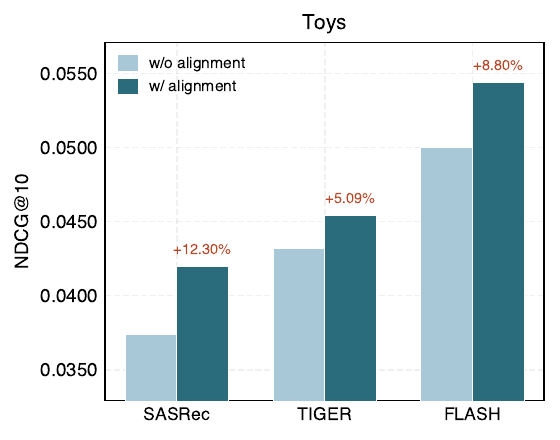}
    \end{subfigure}
    \caption{Effect of semantic alignment across models. ``w/'' and ``w/o'' denote training with and without the alignment loss, respectively.}
    \label{fig:ablation_alignment}
\end{figure}

As shown in Figure~\ref{fig:ablation_alignment}, incorporating the alignment loss consistently improves performance across models and metrics. 
This demonstrates that explicitly anchoring item representations to the original LLM embedding space effectively compensates for the loss of fine-grained semantic information introduced by hashing.
Moreover, this trend generalizes across different models, suggesting that semantic alignment serves as a general and effective mechanism for enhancing representation quality. 
For example, on the Beauty dataset, alignment improves R@10 by $13.06\%$, $9.57\%$, and $7.49\%$ for SASRec, TIGER, and \mymodel, respectively.
Overall, these results highlight the importance of semantic grounding in mitigating information loss introduced by discretization, complementing structural compatibility in determining model effectiveness.
Further qualitative analysis of the learned code embeddings is provided in Appendix~\ref{sec:app_embedding}.

Taken together, these results demonstrate that structural compatibility and semantic grounding are both essential, while decoding capacity further contributes to the remaining performance gap.

\subsection{Hyperparameter Analysis}
\label{sec:hyper}
We analyze the impact of key hyperparameters on the Beauty dataset for clarity. Results on the remaining datasets are provided in Appendix~\ref{sec:app_hparam}, which show similar trends.

\begin{figure}[t!]
    \centering
    \begin{subfigure}{0.32\textwidth}
        \centering
        \includegraphics[width=\linewidth]{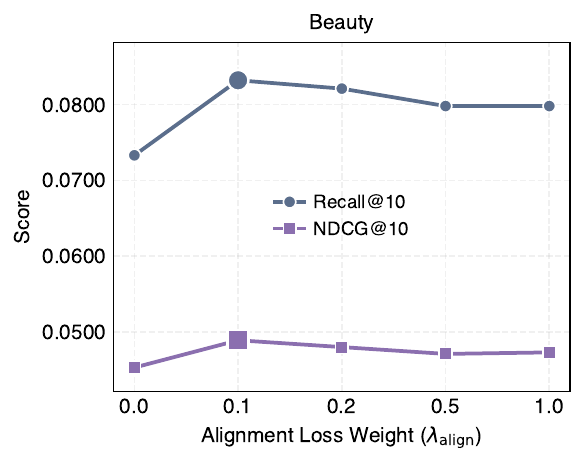}
        \label{fig:hparam-a}
    \end{subfigure}
    \hfill
    \begin{subfigure}{0.32\textwidth}
        \centering
        \includegraphics[width=\linewidth]{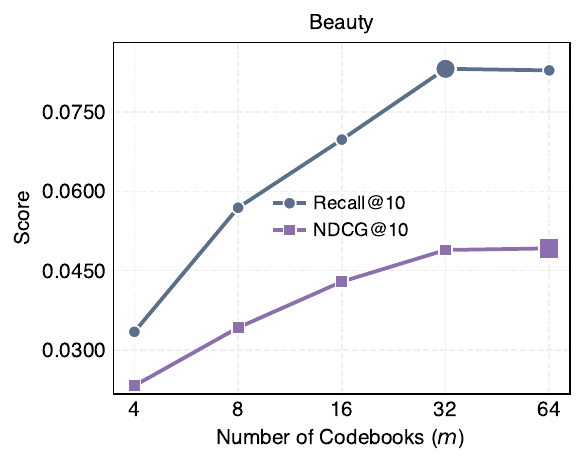}
        \label{fig:hparam-b}
    \end{subfigure}
    \hfill
    \begin{subfigure}{0.32\textwidth}
        \centering
        \includegraphics[width=\linewidth]{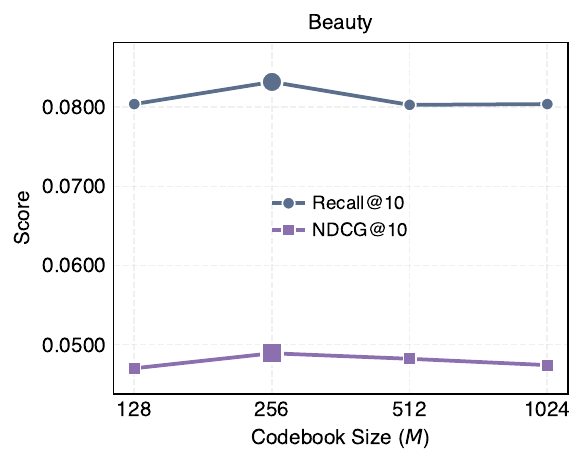}
        \label{fig:hparam-c}
    \end{subfigure}
    \vspace{-15pt}
    \caption{Performance of \mymodel\ over different hyperparameters on Beauty dataset.}
    \label{fig:hparam}
\end{figure}

\textbf{Alignment loss weight $\lambda_{\text{align}}$.}
The alignment weight controls the strength of semantic grounding by balancing the alignment loss against the recommendation objective. We vary $\lambda_{\text{align}}$ in the range~$[0, 1]$.
As shown in Figure~\ref{fig:hparam}~(left), introducing a moderate alignment weight (e.g., 0.1–0.2) consistently improves performance, validating the effectiveness of semantic grounding in compensating for the information loss introduced by discrete tokenization. 
However, excessively large alignment weights (e.g., $\lambda_{\text{align}} \geq 0.5$) fail to provide further improvements and may even lead to performance degradation on other datasets. This suggests that overly strong alignment may dominate the training objective and weaken the contribution of the recommendation signal.

\textbf{Number of codebooks $m$.}
This parameter determines the length of the semantic ID by specifying the number of code positions used to represent each item, thereby affecting the expressiveness of the representation. We experiment with $m \in \{4, 8, 16, 32, 64\}$ while fixing the codebook size.
Figure~\ref{fig:hparam}~(middle) shows that performance improves as $m$ increases, since longer semantic IDs provide greater representational capacity to capture fine-grained item semantics. The gains are substantial when increasing $m$ from 4 to 32, while further increasing $m$ to 64 yields only marginal improvements. 
On other datasets, the performance peak occurs at different values of $m$, indicating that the optimal number of codebooks depends on the dataset scale and interaction sparsity, reflecting a trade-off between representation capacity and modeling complexity.

\textbf{Codebook size $M$.}
The codebook size defines the number of discrete entries available at each code position, controlling the granularity of the semantic representation. We evaluate $M \in \{128, 256, 512, 1024\}$ while fixing the number of codebooks.
From Figure~\ref{fig:hparam}~(right), we observe that increasing $M$ leads to modest performance improvements, as larger codebooks allow for finer-grained distinctions between items. Performance improves from $M=128$ to $M=256$, after which it slightly declines.
This suggests that while larger codebooks increase representational granularity, excessively large $M$ makes learning more difficult due to data sparsity.
Larger codebooks also require higher memory and computational costs, making moderate $M$ more practical.
Compared to increasing the number of codebooks $m$, which directly expands representational capacity through longer semantic ID sequences, the impact of $M$ is relatively limited.
This indicates that capacity is more effectively improved by increasing sequence length rather than per-position granularity.

\subsection{Cold-Start Generalization}
\vspace{-2pt}
To evaluate the generalization ability of \mymodel, we adopt a cold-start setting following TIGER~\cite{tiger}, where a subset of target items is held out from training and used only for evaluation. Full details are provided in Appendix~\ref{sec:app_cold_start}.
From Figure~\ref{fig:cold_start}, we can observe that \mymodel\ consistently outperforms all baselines across all datasets. Compared to SASRec, the performance gap highlights the advantage of semantic ID-based modeling over conventional item ID-based methods in handling unseen items. Compared to TIGER, which employs learned quantization, \mymodel\ still achieves better performance.
This improvement can be attributed to two key factors. 
First, the training-free SimHash tokenization preserves coarse-grained semantic similarity derived from LLM-based text embeddings, enabling stable and meaningful representations for unseen items without relying on interaction data or learned codebooks. 
Second, the semantic alignment objective further anchors item representations to the original embedding space, improving generalization under distribution shift.
Overall, the superior performance of \mymodel\ demonstrates its ability to generalize to unseen items and highlights the importance of semantic grounding for robust recommendation in cold-start scenarios.
\setlength{\intextsep}{6pt}
\setlength{\textfloatsep}{6pt}
\begin{figure}[t!]
    \centering
    \includegraphics[width=0.85\linewidth]{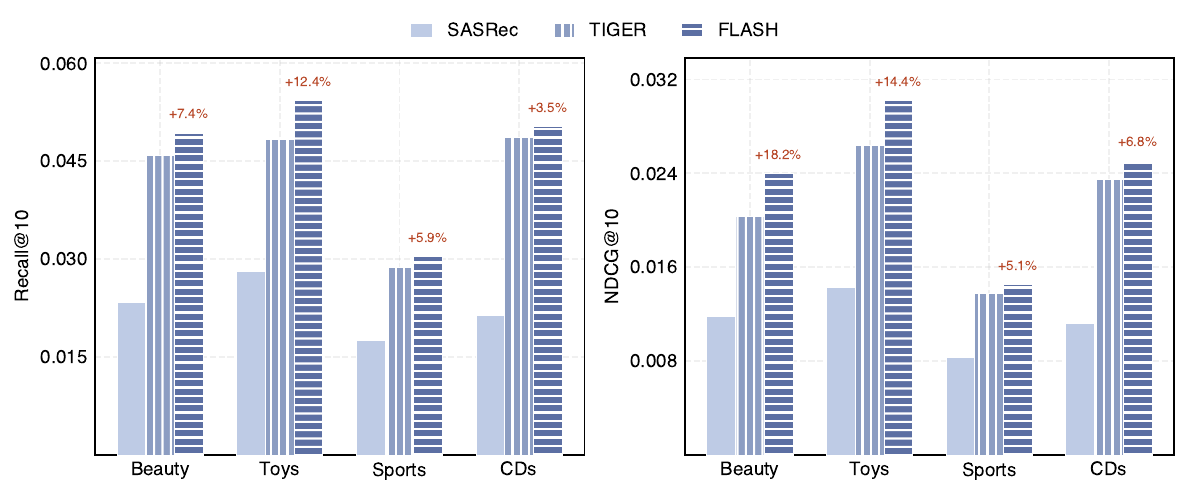}
    \vspace{-5pt}
    \caption{Performance comparison under the cold-start setting, where target items are unseen during training. \mymodel\ achieves the best performance across all datasets. Percentages denote relative improvements of \mymodel\ over TIGER.}
    \label{fig:cold_start}
\end{figure}

\section{Related Work}

\textbf{LLM-based Generative Recommendation.}
Traditional sequential recommendation relies on item ID-based paradigms, where each item is represented as a unique identifier and predicted via embedding matching~\cite{sasrec,gru4rec,bert4rec,caser,hgn,he2020lightgcn,chen2021structured,chang2021sequential,li2022time,chen2022denoising}. Although effective, such approaches lack semantic compositionality and struggle to generalize to new or infrequent items.
With the advancement of large language models (LLMs), generative recommendation has emerged as a promising paradigm that formulates recommendation as sequence generation~\cite{hstu,mtgr,li2025dimerec,lin2025order,liu2024visual,jin2024llm}. Some methods~\cite{p5,collm,tallrec} treat recommendation as a text generation task, leveraging pretrained semantic knowledge. 
However, operating directly in natural-language space without explicit discrete item representations makes efficient large-scale retrieval challenging.
Semantic ID-based methods address this limitation by representing items as discrete token sequences from a shared vocabulary~\cite{tiger,rpg,dou2026transactiongpt}, enabling compositional representations and scalable retrieval.
Despite their success, existing approaches differ significantly in how semantic IDs are constructed and decoded.

\textbf{Semantic ID-based Generative Recommendation.}
Existing approaches to semantic ID construction can be broadly categorized into two dominant quantization-based strategies.
RQ-based methods~\cite{tiger,letter,diger,li2026lsig} produce hierarchically dependent codes typically paired with autoregressive decoding. For example, TIGER~\cite{tiger} employs RQ-VAE, which iteratively quantizes item embedding residuals across multiple codebooks to construct a multi-level tuple of discrete codewords as the item’s semantic ID.
In contrast, PQ-based methods~\cite{rpg, vqrec,recjpq} generate unordered codes compatible with parallel prediction.
Despite these differences, both approaches rely on data-dependent quantizers learned from item embeddings, introducing additional training complexity and potential misalignment with recommendation objectives.
Recent work further explores semantic IDs tailored to bidirectional or diffusion-based decoding~\cite{lladarec,liu2026diffgrm,mdgr}; see Appendix~\ref{sec:app_diffusion} for discussion.
As an alternative to learned quantization, locality-sensitive hashing (LSH) methods~\cite{indyk1997locality,indyk1998approximate,yeh2024rpmixer}, such as SimHash~\cite{charikar2002similarity}, provide a training-free mechanism for constructing semantic IDs via random projections, without requiring quantizer learning.
However, prior work~\cite{tiger} reports inferior performance of hashing-based tokenization, leading to the common belief that such methods are less effective. In this work, we revisit this conclusion by examining the interaction between tokenization structure and decoding strategy in generative recommendation.
\section{Conclusion}
\label{sec:conclusion}
In this work, we demonstrate that the effectiveness of semantic ID-based generative recommendation depends less on tokenization complexity and more on structural compatibility with the decoding paradigm and alignment with the underlying semantic space.
Specifically, our empirical results reveal that a lightweight, training-free SimHash tokenization scheme can achieve competitive performance when paired with parallel decoding and explicit semantic grounding.
It also exhibits stronger generalization than multi-stage learned quantization pipelines.
These findings suggest a critical shift in perspective for generative recommendation: 
cross-stage compatibility and semantic alignment may be more important than increasing tokenization complexity. This highlights the potential of simple, principled designs to rival or even outperform more complex alternatives.

\paragraph{Limitations.}
Despite its strong performance and efficiency, \mymodel\ has several limitations. 
First, it relies on pre-computed text embeddings, so its effectiveness depends on the quality of the underlying representations and may degrade when textual information is sparse or noisy.
Second, hashing-based tokenization may introduce collisions.
Although our empirical analysis shows that collisions are limited under the evaluated configurations, they may become more pronounced at larger scales or under different codebook settings, potentially reducing the discriminability of semantic IDs.
Third, our experiments are conducted on public recommendation benchmarks, and industrial-scale validation remains necessary to assess the scalability and robustness of \mymodel\ in real-world recommendation systems.
Finally, our study focuses on textual signals and does not explicitly incorporate other modalities such as images. Extending the framework to richer multimodal settings is an important direction for future work, particularly when item semantics cannot be fully captured by text alone.

\begin{ack}
This work is supported in part by NSF under grants III-2106758 and POSE-2346158. We thank the anonymous reviewers of NeurIPS 2026 for their valuable feedback and suggestions.
\end{ack}

\bibliographystyle{plain} 
\bibliography{references}

@String{Computing = "Computing" }

@String{Computer = "{IEEE} Computer" }

@article{tiger,
  title={{Recommender Systems with Generative Retrieval}},
  author={Rajput, Shashank and Mehta, Nikhil and Singh, Anima and Hulikal Keshavan, Raghunandan and Vu, Trung and Heldt, Lukasz and Hong, Lichan and Tay, Yi and Tran, Vinh and Samost, Jonah and others},
  journal={Advances in Neural Information Processing Systems},
  volume={36},
  pages={10299--10315},
  year={2023}
}

@inproceedings{rpg,
  title={{Generating Long Semantic {IDs} in Parallel for Recommendation}},
  author={Hou, Yupeng and Li, Jiacheng and Shin, Ashley and Jeon, Jinsung and Santhanam, Abhishek and Shao, Wei and Hassani, Kaveh and Yao, Ning and McAuley, Julian},
  booktitle={Proceedings of the 31st ACM SIGKDD Conference on Knowledge Discovery and Data Mining},
  pages={956--966},
  year={2025}
}

@inproceedings{vqrec,
  title={{Learning Vector-Quantized Item Representation for Transferable Sequential Recommenders}},
  author={Hou, Yupeng and He, Zhankui and McAuley, Julian and Zhao, Wayne Xin},
  booktitle={Proceedings of the ACM Web Conference 2023},
  pages={1162--1171},
  year={2023}
}

@inproceedings{letter,
  title={{Learnable Item Tokenization for Generative Recommendation}},
  author={Wang, Wenjie and Bao, Honghui and Lin, Xinyu and Zhang, Jizhi and Li, Yongqi and Feng, Fuli and Ng, See-Kiong and Chua, Tat-Seng},
  booktitle={Proceedings of the 33rd ACM International Conference on Information and Knowledge Management},
  pages={2400--2409},
  year={2024}
}

@inproceedings{recjpq,
  title={{{RecJPQ}: Training Large-Catalogue Sequential Recommenders}},
  author={Petrov, Aleksandr V and Macdonald, Craig},
  booktitle={Proceedings of the 17th ACM International Conference on Web Search and Data Mining},
  pages={538--547},
  year={2024}
}

@inproceedings{diger,
  title={{Differentiable Semantic {ID} for Generative Recommendation}},
  author={Fu, Junchen and Ge, Xuri and Karatzoglou, Alexandros and Arapakis, Ioannis and Verberne, Suzan and Jose, Joemon M and Ren, Zhaochun},
booktitle = {Proceedings of the 49th International ACM SIGIR Conference on Research and Development in Information Retrieval},
  year={2026},
  pages={369--379}
}

@inproceedings{indyk1997locality,
  title={{Locality-Preserving Hashing in Multidimensional Spaces}},
  author={Indyk, Piotr and Motwani, Rajeev and Raghavan, Prabhakar and Vempala, Santosh},
  booktitle={Proceedings of the Twenty-Ninth Annual ACM Symposium on Theory of Computing},
  pages={618--625},
  year={1997}
}

@inproceedings{indyk1998approximate,
  title={{Approximate Nearest Neighbors: Towards Removing the Curse of Dimensionality}},
  author={Indyk, Piotr and Motwani, Rajeev},
  booktitle={Proceedings of the Thirtieth Annual ACM Symposium on Theory of Computing},
  pages={604--613},
  year={1998}
}

@inproceedings{charikar2002similarity,
  title={{Similarity Estimation Techniques from Rounding Algorithms}},
  author={Charikar, Moses S},
  booktitle={Proceedings of the Thirty-Fourth Annual ACM Symposium on Theory of Computing},
  pages={380--388},
  year={2002}
}

@inproceedings{caser,
  title={{Personalized Top-{N} Sequential Recommendation via Convolutional Sequence Embedding}},
  author={Tang, Jiaxi and Wang, Ke},
  booktitle={Proceedings of the Eleventh ACM International Conference on Web Search and Data Mining},
  pages={565--573},
  year={2018}
}

@inproceedings{gru4rec,
  title={{Session-Based Recommendations with Recurrent Neural Networks}},
  author={Hidasi, Bal{\'a}zs and Karatzoglou, Alexandros and Baltrunas, Linas and Tikk, Domonkos},
  booktitle={International Conference on Learning Representations},
  year={2016}
}

@inproceedings{hgn,
  title={{Hierarchical Gating Networks for Sequential Recommendation}},
  author={Ma, Chen and Kang, Peng and Liu, Xue},
  booktitle={Proceedings of the 25th ACM SIGKDD International Conference on Knowledge Discovery \& Data Mining},
  pages={825--833},
  year={2019}
}

@inproceedings{bert4rec,
  title={{{BERT4Rec}: Sequential Recommendation with Bidirectional Encoder Representations from {Transformer}}},
  author={Sun, Fei and Liu, Jun and Wu, Jian and Pei, Changhua and Lin, Xiao and Ou, Wenwu and Jiang, Peng},
  booktitle={Proceedings of the 28th ACM International Conference on Information and Knowledge Management},
  pages={1441--1450},
  year={2019}
}

@inproceedings{sasrec,
  title={{Self-Attentive Sequential Recommendation}},
  author={Kang, Wang-Cheng and McAuley, Julian},
  booktitle={Proceedings of IEEE International Conference on Data Mining},
  pages={197--206},
  year={2018}
}

@inproceedings{s3rec,
  title={{{S$^3$-Rec}: Self-Supervised Learning for Sequential Recommendation with Mutual Information Maximization}},
  author={Zhou, Kun and Wang, Hui and Zhao, Wayne Xin and Zhu, Yutao and Wang, Sirui and Zhang, Fuzheng and Wang, Zhongyuan and Wen, Ji-Rong},
  booktitle={Proceedings of the 29th ACM International Conference on Information \& Knowledge Management},
  pages={1893--1902},
  year={2020}
}

@inproceedings{dataset,
  title={{Image-Based Recommendations on Styles and Substitutes}},
  author={McAuley, Julian and Targett, Christopher and Shi, Qinfeng and Van Den Hengel, Anton},
  booktitle={Proceedings of the 38th International ACM SIGIR Conference on Research and Development in Information Retrieval},
  pages={43--52},
  year={2015}
}

@inproceedings{p5,
  title={{Recommendation as Language Processing ({RLP}): A Unified Pretrain, Personalized Prompt \& Predict Paradigm ({P5})}},
  author={Geng, Shijie and Liu, Shuchang and Fu, Zuohui and Ge, Yingqiang and Zhang, Yongfeng},
  booktitle={Proceedings of the 16th ACM Conference on Recommender Systems},
  pages={299--315},
  year={2022}
}

@article{collm,
  title={{{CoLLM}: Integrating Collaborative Embeddings into Large Language Models for Recommendation}},
  author={Zhang, Yang and Feng, Fuli and Zhang, Jizhi and Bao, Keqin and Wang, Qifan and He, Xiangnan},
  journal={IEEE Transactions on Knowledge and Data Engineering},
  year={2025},
  publisher={IEEE},
  volume={37},
  number={5},
  pages={2329--2340}
}

@inproceedings{tallrec,
  title={{{TALLRec}: An Effective and Efficient Tuning Framework to Align Large Language Model with Recommendation}},
  author={Bao, Keqin and Zhang, Jizhi and Zhang, Yang and Wang, Wenjie and Feng, Fuli and He, Xiangnan},
  booktitle={Proceedings of the 17th ACM Conference on Recommender Systems},
  pages={1007--1014},
  year={2023}
}

@inproceedings{hstu,
  title={{Actions Speak Louder Than Words: Trillion-Parameter Sequential Transducers for Generative Recommendations}},
  author={Zhai, Jiaqi and Liao, Lucy and Liu, Xing and Wang, Yueming and Li, Rui and Cao, Xuan and Gao, Leon and Gong, Zhaojie and Gu, Fangda and He, Jiayuan and others},
booktitle = {Proceedings of the 41st International Conference on Machine Learning},
  year={2024},
  pages={58484--58509}
}

@inproceedings{chen2021structured,
  title={{Structured Graph Convolutional Networks with Stochastic Masks for Recommender Systems}},
  author={Chen, Huiyuan and Wang, Lan and Lin, Yusan and Yeh, Chin-Chia Michael and Wang, Fei and Yang, Hao},
  booktitle={Proceedings of the 44th International ACM SIGIR Conference on Research and Development in Information Retrieval},
  pages={614--623},
  year={2021}
}

@inproceedings{mtgr,
  title={{{MTGR}: Industrial-Scale Generative Recommendation Framework in {Meituan}}},
  author={Han, Ruidong and Yin, Bin and Chen, Shangyu and Jiang, He and Jiang, Fei and Li, Xiang and Ma, Chi and Huang, Mincong and Li, Xiaoguang and Jing, Chunzhen and others},
  booktitle={Proceedings of the 34th ACM International Conference on Information and Knowledge Management},
  pages={5731--5738},
  year={2025}
}

@inproceedings{chen2022denoising,
  title={{Denoising Self-Attentive Sequential Recommendation}},
  author={Chen, Huiyuan and Lin, Yusan and Pan, Menghai and Wang, Lan and Yeh, Chin-Chia Michael and Li, Xiaoting and Zheng, Yan and Wang, Fei and Yang, Hao},
  booktitle={Proceedings of the 16th ACM Conference on Recommender Systems},
  pages={92--101},
  year={2022}
}

@inproceedings{li2025dimerec,
  title={{{DimeRec}: A Unified Framework for Enhanced Sequential Recommendation via Generative Diffusion Models}},
  author={Li, Wuchao and Huang, Rui and Zhao, Haijun and Liu, Chi and Zheng, Kai and Liu, Qi and Mou, Na and Zhou, Guorui and Lian, Defu and Song, Yang and others},
  booktitle={Proceedings of the Eighteenth ACM International Conference on Web Search and Data Mining},
  pages={726--734},
  year={2025}
}

@inproceedings{lin2025order,
  title={{Order-Agnostic Identifier for Large Language Model-Based Generative Recommendation}},
  author={Lin, Xinyu and Shi, Haihan and Wang, Wenjie and Feng, Fuli and Wang, Qifan and Ng, See-Kiong and Chua, Tat-Seng},
  booktitle={Proceedings of the 48th International ACM SIGIR Conference on Research and Development in Information Retrieval},
  pages={1923--1933},
  year={2025}
}

@inproceedings{he2020lightgcn,
  title={{{LightGCN}: Simplifying and Powering Graph Convolution Network for Recommendation}},
  author={He, Xiangnan and Deng, Kuan and Wang, Xiang and Li, Yan and Zhang, Yongdong and Wang, Meng},
  booktitle={Proceedings of the 43rd International ACM SIGIR Conference on Research and Development in Information Retrieval},
  pages={639--648},
  year={2020}
}

@inproceedings{chang2021sequential,
  title={{Sequential Recommendation with Graph Neural Networks}},
  author={Chang, Jianxin and Gao, Chen and Zheng, Yu and Hui, Yiqun and Niu, Yanan and Song, Yang and Jin, Depeng and Li, Yong},
  booktitle={Proceedings of the 44th International ACM SIGIR Conference on Research and Development in Information Retrieval},
  pages={378--387},
  year={2021}
}

@inproceedings{li2026lsig,
  title={{{LSIG}: Long Semantic {IDs} for Generative Recommendation}},
  author={Li, Zhao and Qi, FengYang and Xu, Chuanyu and Zhang, Tao and Huo, Chengfu and Zhang, Peng},
  booktitle={Proceedings of the ACM Web Conference 2026},
  pages={7779--7788},
  year={2026}
}

@inproceedings{chen2020simple,
  title={{A Simple Framework for Contrastive Learning of Visual Representations}},
  author={Chen, Ting and Kornblith, Simon and Norouzi, Mohammad and Hinton, Geoffrey},
  booktitle={Proceedings of the 37th International Conference on Machine Learning},
  pages={1597--1607},
  year={2020},
  organization={PMLR}
}

@inproceedings{cai2024medusa,
  title={{{MEDUSA}: Simple {LLM} Inference Acceleration Framework with Multiple Decoding Heads}},
  author={Cai, Tianle and Li, Yuhong and Geng, Zhengyang and Peng, Hongwu and Lee, Jason D and Chen, Deming and Dao, Tri},
  booktitle={Proceedings of the 41st International Conference on Machine Learning},
  year={2024},
  pages={5209--5235}
}

@inproceedings{yeh2024rpmixer,
  title={{{RPMixer}: Shaking Up Time Series Forecasting with Random Projections for Large Spatial-Temporal Data}},
  author={Yeh, Chin-Chia Michael and Fan, Yujie and Dai, Xin and Saini, Uday Singh and Lai, Vivian and Aboagye, Prince Osei and Wang, Junpeng and Chen, Huiyuan and Zheng, Yan and Zhuang, Zhongfang and others},
  booktitle={Proceedings of the 30th ACM SIGKDD Conference on Knowledge Discovery and Data Mining},
  pages={3919--3930},
  year={2024}
}

@inproceedings{chen2024masked,
  title={{Masked Graph Transformer for Large-Scale Recommendation}},
  author={Chen, Huiyuan and Xu, Zhe and Yeh, Chin-Chia Michael and Lai, Vivian and Zheng, Yan and Xu, Minghua and Tong, Hanghang},
  booktitle={Proceedings of the 47th International ACM SIGIR Conference on Research and Development in Information Retrieval},
  pages={2502--2506},
  year={2024}
}

@inproceedings{zheng2024adapting,
  title={{Adapting Large Language Models by Integrating Collaborative Semantics for Recommendation}},
  author={Zheng, Bowen and Hou, Yupeng and Lu, Hongyu and Chen, Yu and Zhao, Wayne Xin and Chen, Ming and Wen, Ji-Rong},
  booktitle={IEEE 40th International Conference on Data Engineering},
  pages={1435--1448},
  year={2024},
  organization={IEEE}
}

@inproceedings{dou2026transactiongpt,
  title={{{TransactionGPT}: Toward Foundational Transaction Modeling}},
  author={Dou, Yingtong and Jiang, Zhimeng and Zhang, Tianyi and Hu, Mingzhi and Xu, Zhichao and Chen, Huiyuan and Jain, Shubham and Saini, Uday Singh and Fan, Xiran and Sun, Jiarui and others},
  booktitle={Proceedings of the 32nd ACM SIGKDD Conference on Knowledge Discovery and Data Mining},
  pages={7185--7196},
  year={2026}
}

@inproceedings{
jin2024llm,
title={{{LLM} Maybe {LongLM}: {SelfExtend} {LLM} Context Window Without Tuning}},
author={Hongye Jin and Xiaotian Han and Jingfeng Yang and Zhimeng Jiang and Zirui Liu and Chia-Yuan Chang and Huiyuan Chen and Xia Hu},
booktitle={Proceedings of the 41st International Conference on Machine Learning},
pages={22099--22114},
year={2024}
}

@inproceedings{li2024large,
  title={{Large Language Models for Generative Recommendation: A Survey and Visionary Discussions}},
  author={Li, Lei and Zhang, Yongfeng and Liu, Dugang and Chen, Li},
  booktitle={Proceedings of the 2024 Joint International Conference on Computational Linguistics, Language Resources and Evaluation (LREC-COLING 2024)},
  pages={10146--10159},
  year={2024}
}

@inproceedings{rendle2012bpr,
  title={{{BPR}: {Bayesian} Personalized Ranking from Implicit Feedback}},
  author={Rendle, Steffen and Freudenthaler, Christoph and Gantner, Zeno and Schmidt-Thieme, Lars},
  booktitle = {Proceedings of the Twenty-Fifth Conference on Uncertainty in Artificial Intelligence},
  year={2009},
  pages={452--461}
}

@inproceedings{wang2024eager,
  title={{{EAGER}: Two-Stream Generative Recommender with Behavior-Semantic Collaboration}},
  author={Wang, Ye and Xun, Jiahao and Hong, Minjie and Zhu, Jieming and Jin, Tao and Lin, Wang and Li, Haoyuan and Li, Linjun and Xia, Yan and Zhao, Zhou and others},
  booktitle={Proceedings of the 30th ACM SIGKDD Conference on Knowledge Discovery and Data Mining},
  pages={3245--3254},
  year={2024}
}

@article{ge2013optimized,
  title={{Optimized Product Quantization}},
  author={Ge, Tiezheng and He, Kaiming and Ke, Qifa and Sun, Jian},
  journal={IEEE Transactions on Pattern Analysis and Machine Intelligence},
  volume={36},
  number={4},
  pages={744--755},
  year={2014},
  publisher={IEEE}
}

@inproceedings{hua2023index,
  title={{How to Index Item {IDs} for Recommendation Foundation Models}},
  author={Hua, Wenyue and Xu, Shuyuan and Ge, Yingqiang and Zhang, Yongfeng},
  booktitle={Proceedings of the Annual International ACM SIGIR Conference on Research and Development in Information Retrieval in the Asia Pacific Region},
  pages={195--204},
  year={2023}
}

@article{jegou2010product,
  title={{Product Quantization for Nearest Neighbor Search}},
  author={J{\'e}gou, Herv{\'e} and Douze, Matthijs and Schmid, Cordelia},
  journal={IEEE Transactions on Pattern Analysis and Machine Intelligence},
  volume={33},
  number={1},
  pages={117--128},
  year={2011},
  publisher={IEEE}
}

@inproceedings{lee2022autoregressive,
  title={{Autoregressive Image Generation Using Residual Quantization}},
  author={Lee, Doyup and Kim, Chiheon and Kim, Saehoon and Cho, Minsu and Han, Wook-Shin},
  booktitle={Proceedings of the IEEE/CVF Conference on Computer Vision and Pattern Recognition},
  pages={11513--11522},
  year={2022}
}

@article{van2017neural,
  title={{Neural Discrete Representation Learning}},
  author={van den Oord, Aaron and Vinyals, Oriol and Koray Kavukcuoglu},
  journal={Advances in Neural Information Processing Systems},
  pages={6309--6318},
  year={2017}
}

@article{lladarec,
  title={{{LLaDA-Rec}: Discrete Diffusion for Parallel Semantic {ID} Generation in Generative Recommendation}},
  author={Shi, Teng and Shen, Chenglei and Yu, Weijie and Nie, Shen and Li, Chongxuan and Zhang, Xiao and He, Ming and Han, Yan and Xu, Jun},
  journal={arXiv preprint arXiv:2511.06254},
  year={2025}
}

@inproceedings{liu2026diffgrm,
  title={{{DiffGRM}: Diffusion-Based Generative Recommendation Model}},
  author={Liu, Zhao and Zhu, Yichen and Yang, Yiqing and Lv, Xiao and Tang, Guoping and Huang, Rui and Luo, Qiang and Tang, Ruiming and Zhou, Guorui},
  booktitle={Proceedings of the ACM Web Conference 2026},
  pages={5853--5864},
  year={2026}
}

@article{mdgr,
  title={{Masked Diffusion Generative Recommendation}},
  author={Mu, Lingyu and Deng, Hao and Xing, Haibo and Hu, Jinxin and Zhang, Yu and Zeng, Xiaoyi and Zhang, Jing},
  journal={arXiv preprint arXiv:2601.19501},
  year={2026}
}

@inproceedings{li2022time,
  title={{Time-Aware Hyperbolic Graph Attention Network for Session-Based Recommendation}},
  author={Li, Xiaohan and Liu, Yuqing and Liu, Zheng and Yu, Philip S},
  booktitle={2022 IEEE International Conference on Big Data (Big Data)},
  pages={626--635},
  year={2022},
  organization={IEEE}
}

@inproceedings{liu2024visual,
  title={{{Visual Summary Thought} of Large Vision-Language Models for Multimodal Recommendation}},
  author={Liu, Yuqing and Wang, Yu and Cao, Yuwei and Sun, Lichao and Yu, Philip S},
  booktitle={2024 IEEE International Conference on Big Data (BigData)},
  pages={456--461},
  year={2024},
  organization={IEEE}
}

\newpage
\appendix
\section{Additional Experimental Details}
\label{sec:app_details}
\subsection{Dataset Statistics and Preprocessing}
\label{sec:app_dataset}
We conduct experiments on four widely used categories from the Amazon Product Reviews dataset~\cite{dataset}.
Following prior work~\cite{tiger,rpg}, we construct user interaction sequences by treating each review as an interaction and sorting them chronologically. We filter out users and items with fewer than 5 interactions to ensure sufficient behavioral signals. We adopt the widely used leave-last-out evaluation protocol~\cite{sasrec,p5,zheng2024adapting}, where the last item in each sequence is used for testing, the second-to-last item for validation, and the remaining interactions for training. Table~\ref{tab:datasets} summarizes the statistics of the processed datasets.

\begin{table}[ht]
  \centering
  \caption{Statistics of the processed datasets.}
  \label{tab:datasets}
  \small
  \setlength{\tabcolsep}{4pt}
  \begin{tabular}{lcccc}
    \toprule
    Datasets
    & \makecell{\# Users}
    & \makecell{\# Items}
    & \makecell{\# Interactions}
    & \makecell{Average \\ Sequence Length} \\
    \midrule
    Beauty & 22,363 & 12,101 &    198,502  &  8.87 \\
    Toys & 19,412 & 11,924 &    167,597 &  8.63 \\
    Sports & 35,598 & 18,357 &    296,337 &  8.32 \\
    CDs    & 75,258 & 64,443 & 1,097,592  & 14.58 \\
    \bottomrule
  \end{tabular}
\end{table}

\subsection{Baselines}
\label{sec:app_baselines}

We compare \mymodel\ against three categories of baselines:

\noindent\textbf{(1) Item ID-based methods.}
These methods represent each item using a unique identifier and learn item embeddings solely from interaction data.
\textbf{Caser}~\cite{caser} models user–item interaction sequences using convolutional neural networks to capture both union-level and point-level sequential patterns.
\textbf{GRU4Rec}~\cite{gru4rec} utilizes recurrent neural networks to model sequential dependencies in user behavior.
\textbf{HGN}~\cite{hgn} introduces a hierarchical gating mechanism to adaptively combine long-term user preferences with short-term sequential signals.
\textbf{BERT4Rec}~\cite{bert4rec} applies bidirectional Transformer encoders with a masked item prediction objective to capture contextual dependencies in interaction sequences.
\textbf{SASRec}~\cite{sasrec} employs a self-attention mechanism to model user sequences, enabling efficient capture of long-range dependencies with a unidirectional Transformer.
\textbf{S$^3$-Rec}~\cite{s3rec} enhances a SASRec-style Transformer with self-supervised pretraining objectives.

\noindent\textbf{(2) RQ-based semantic ID methods.}
These methods construct semantic IDs via residual quantization and typically rely on autoregressive decoding.
\textbf{TIGER}~\cite{tiger} adopts an RQ-VAE~\cite{lee2022autoregressive} to produce hierarchical semantic IDs and models them with autoregressive decoding, generating tokens sequentially with beam search.
\textbf{LETTER}~\cite{letter} learns semantic IDs via residual quantization with additional semantic, collaborative, and diversity constraints to improve token quality.
\textbf{DIGER}~\cite{diger} introduces differentiable residual quantization to jointly optimize tokenization and recommendation in an end-to-end framework.

\noindent\textbf{(3) PQ-based semantic ID methods.}
These methods derive semantic IDs using product quantization, producing independent codes that enable flexible modeling.
\textbf{RecJPQ}~\cite{recjpq} applies joint product quantization to represent items as combinations of shared sub-embeddings.
\textbf{VQ-Rec}~\cite{vqrec} learns vector-quantized item representations via codebooks and integrates them into sequential recommendation models.
\textbf{RPG}~\cite{rpg} generates long semantic IDs via optimized product quantization (OPQ)~\cite{ge2013optimized} and models them with parallel decoding to enable efficient and scalable inference.

\subsection{Implementation Details}
\label{sec:app_implement}
For all experiments, we use OpenAI's \texttt{text-embedding-3-large}\footnote{\label{fn:openai}\url{https://openai.com/index/new-embedding-models-and-api-updates/}} model to extract semantic embeddings for each item. 
Specifically, we construct descriptive sentences from item metadata (e.g., title, description, category), which are encoded into 3072-dimensional embeddings. 
These embeddings are used as inputs to different quantization methods (e.g., RQ-VAE, OPQ, LSH) to ensure a unified input representation, where SimHash is adopted as the specific LSH instantiation.

For the preliminary study illustrated in Figure~\ref{fig:prelim}, we follow TIGER's~\cite{tiger} default settings, using T5 as the sequence backbone with $4$ codebooks, a codebook size of $256$, and a maximum sequence length of $50$. 
For the main experiments, we follow the original backbone configurations of the baselines, while FLASH employs a 2-layer, 2-head Transformer decoder with an embedding dimension $d \in \{448, 896\}$ and a feed-forward network dimension of $1024$.

For the semantic grounding analysis in Section \ref{sec:ablation_align} (Figure~\ref{fig:ablation_alignment}), we apply different alignment strategies depending on the model architecture. 
For each model, the variants with and without semantic alignment use the same backbone and training configuration, differing only in whether the alignment objective is applied.
For SASRec, which does not use a codebook, we directly align its item embeddings with the semantic embeddings. 
For TIGER, we introduce a linear projection layer to map the aggregated codebook representations to the semantic embedding space. This design encourages lower-level representations to capture semantic alignment, while higher layers focus on modeling item–item dependencies within the sequence.  
We keep the original architectures of SASRec and TIGER unchanged, and tune the hyperparameter $\lambda_{\text{align}} \in [0, 1]$ with a step size of $0.1$. 

All experiments are conducted on a single NVIDIA A100 (80GB) GPU. The best hyperparameters of \mymodel\ are summarized in Table~\ref{tab:best_hyperparameters}.

\begin{table}[h]
\centering
\caption{Best hyperparameters for \mymodel\ on each dataset.}
\label{tab:best_hyperparameters}
\small
\setlength{\tabcolsep}{4pt}
\begin{tabular}{lccccccc}
\toprule
Dataset 
& \makecell{\# Layers} 
& \makecell{\# Codebooks \\ ($m$)} 
& \makecell{Codebook \\ Size ($M$)} 
& \makecell{Embedding \\ Dim ($d$)} 
& \makecell{Codebook \\ Dropout} 
& \makecell{Learning \\ Rate} 
& \makecell{Alignment \\ Weight ($\lambda_{\text{align}}$)} \\
\midrule
Beauty  & 2 & 32 & 256 & 448 & 0.2 & 0.01 & 0.1 \\
Toys    & 2 & 64 & 256 & 448 & 0.2 & 0.003 & 0.2 \\
Sports  & 2 & 64 & 256 & 896 & 0.2 & 0.003 & 0.2 \\
CDs     & 2 & 64 & 512 & 448 & 0.2 & 0.001 & 0.1 \\
\bottomrule
\end{tabular}
\end{table}

\subsection{Cold-Start Evaluation Details}
\label{sec:app_cold_start}
To construct the cold-start setting, we follow TIGER's evaluation protocol~\cite{tiger}. Specifically, we randomly sample 5\% of target items (i.e., the last item in each test sequence) to form a cold-start item set. 
We then remove all interactions involving these items from the training set, ensuring that they remain completely unseen during training. 
Finally, we evaluate model performance on the subset of test sequences whose target items belong to this set, thereby simulating a strict cold-start scenario.
For item ID-based baselines such as SASRec~\cite{sasrec}, embeddings of unseen items are randomly initialized at evaluation time, as they are not updated during training.

\section{Additional Experimental Results}
\subsection{Further Analysis of Training Efficiency}
\label{sec:app_efficiency}
We follow the standard configurations for each method. 
TIGER adopts residual quantization with a small number of codebooks (e.g., 4 tokens per item), as commonly used due to the computational cost of autoregressive decoding. 
In contrast, \mymodel\ employs a larger number of codebooks (e.g., 32 tokens per item) enabled by parallel prediction, allowing more expressive semantic IDs. 
SASRec operates on single-token item IDs without semantic tokenization.

\begin{figure}[th]
    \centering
    \begin{subfigure}{0.24\textwidth}
        \centering
        \includegraphics[width=\linewidth]{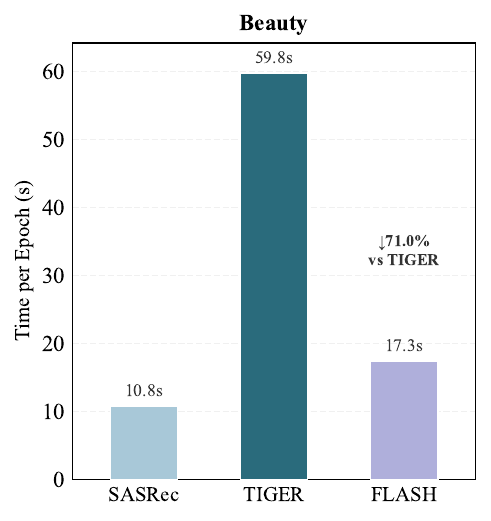}
    \end{subfigure}
    \hfill
    \begin{subfigure}{0.24\textwidth}
        \centering
        \includegraphics[width=\linewidth]{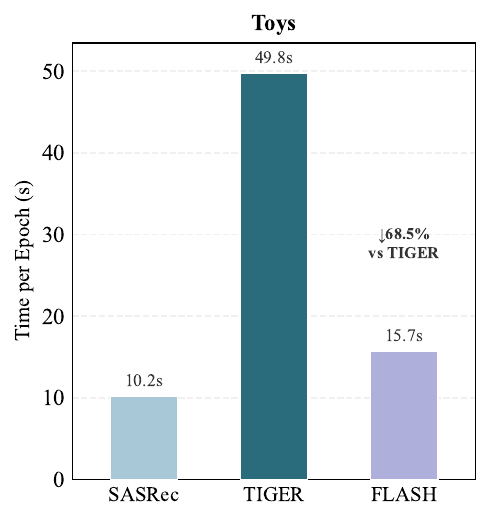}
    \end{subfigure}
    \hfill
    \begin{subfigure}{0.24\textwidth}
        \centering
        \includegraphics[width=\linewidth]{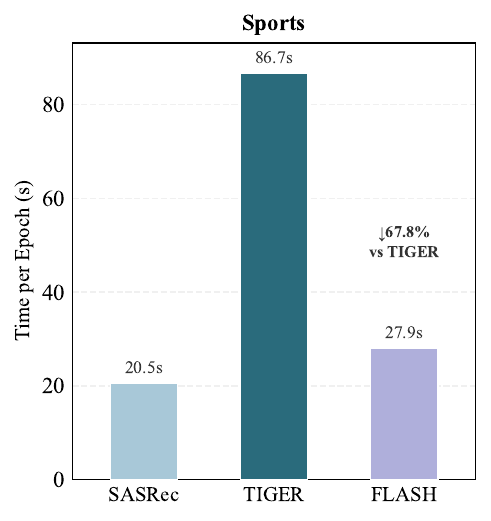}
    \end{subfigure}
     \hfill
    \begin{subfigure}{0.24\textwidth}
        \centering
        \includegraphics[width=\linewidth]{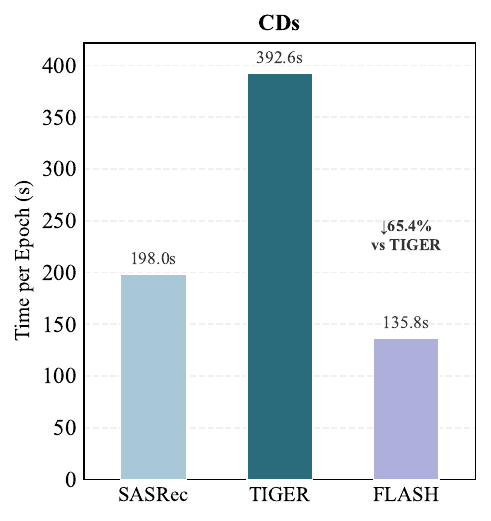}
    \end{subfigure}
    \vspace{-5pt}
    \caption{Average training time per epoch (seconds).}
    \label{fig:app_efficiency}
\end{figure}

We report the average training time per epoch for representative item ID-based and semantic ID-based models in Figure~\ref{fig:app_efficiency}. All models are trained under the same hardware environment and data processing pipeline for a fair comparison.
The reported training time measures the cost of training the sequential recommendation model (Stage II) after semantic IDs are constructed, and does not include the tokenization stage (Stage I). This allows us to isolate the efficiency of the generative recommendation model under different tokenization schemes.
 
\mymodel\ consistently requires substantially less training time than TIGER across all datasets. This improvement is primarily due to the use of parallel prediction, which avoids the sequential dependency inherent in autoregressive decoding and leads to more efficient training. 
Notably, \mymodel\ also employs a training-free tokenization scheme, eliminating the need for an additional quantizer training stage in the overall pipeline.
Together with the tokenization efficiency comparison in Figure~\ref{fig:collision}, these results provide a more complete view of the computational efficiency of the overall pipeline.
Compared to SASRec, \mymodel\ introduces moderate computational overhead, reflecting the increased modeling capacity of semantic ID-based representations.

\subsection{Analysis of Semantic Grounding on Additional Datasets}
\label{sec:app_alignment}
\begin{figure}[htbp]
    \centering
    \begin{subfigure}{0.24\textwidth}
        \centering
        \includegraphics[width=\linewidth]{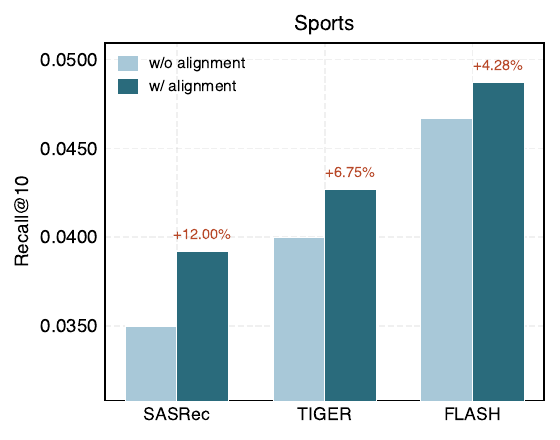}
    \end{subfigure}
    \hfill
    \begin{subfigure}{0.24\textwidth}
        \centering
        \includegraphics[width=\linewidth]{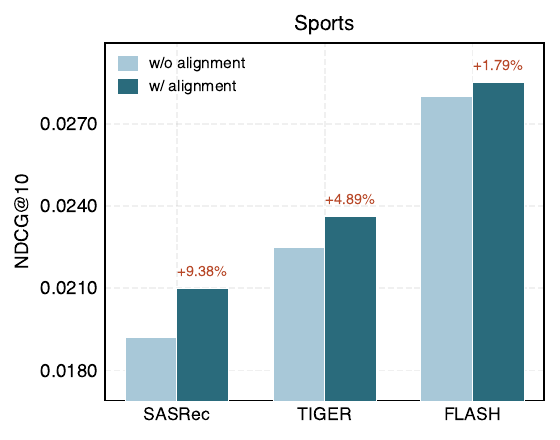}
    \end{subfigure}
    \hfill
    \begin{subfigure}{0.24\textwidth}
        \centering
           \includegraphics[width=\linewidth]{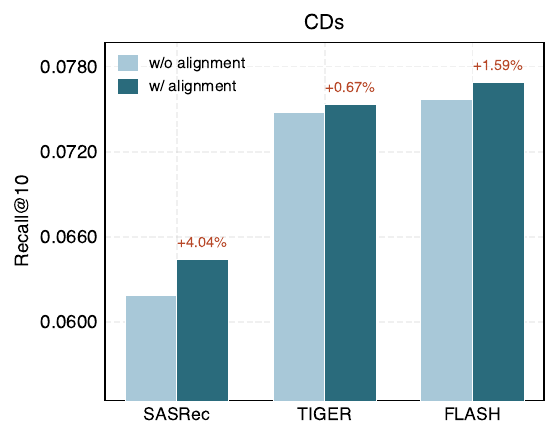}
    \end{subfigure}
    \hfill
    \begin{subfigure}{0.24\textwidth}
        \centering
        \includegraphics[width=\linewidth]{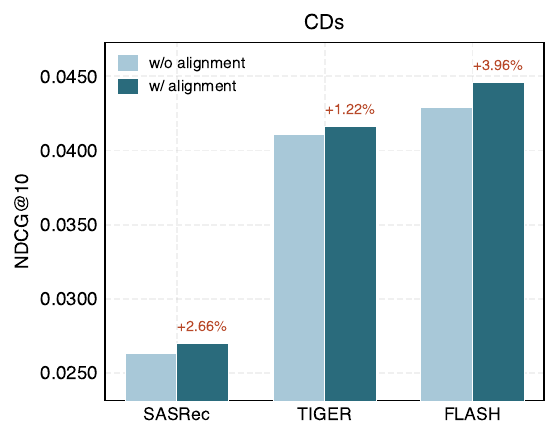}
    \end{subfigure}
    \vspace{-5pt}
    \caption{Effect of semantic alignment on additional datasets. ``w/'' and ``w/o'' denote training with and without the alignment loss, respectively.}
    \label{fig:app_ablation_alignment}
\end{figure}

To complement the analysis in the main text, we report the effect of the alignment objective on the remaining datasets in Figure~\ref{fig:app_ablation_alignment}. Consistent with the observations in Figure~\ref{fig:ablation_alignment}, incorporating the alignment loss leads to improved performance across all evaluated datasets. This consistency suggests that the benefit of semantic grounding is robust across different data distributions, further supporting its role as a general regularization mechanism.

\subsection{Qualitative Analysis of Code Embeddings}
\label{sec:app_embedding}
\begin{figure}[thb]
    \centering
    \includegraphics[width=0.8\linewidth]{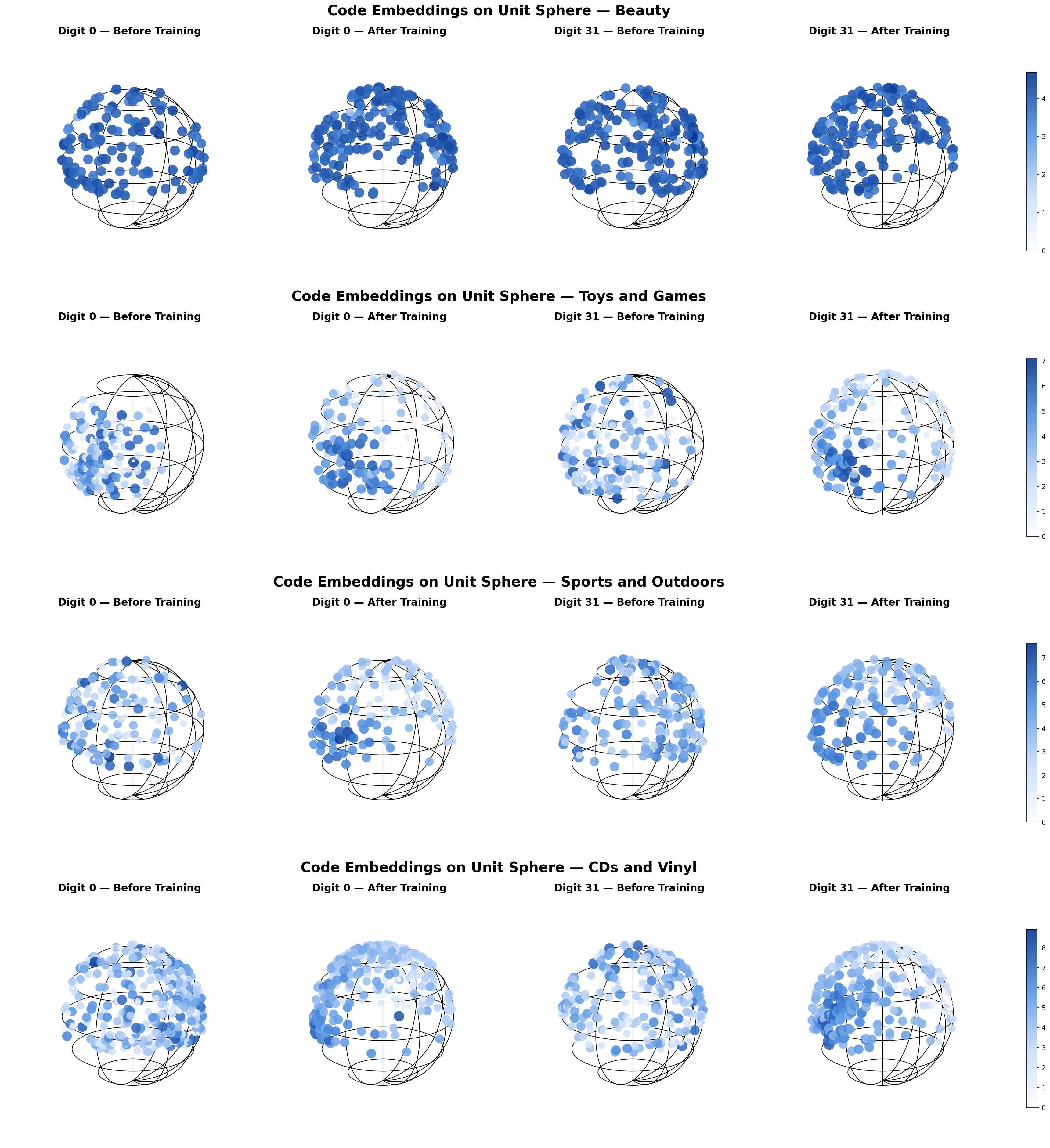}
    \caption{Visualization of learned code embeddings on the unit sphere across different datasets, before and after training. Each subfigure corresponds to a code position (labeled as “Digit” in the figure). 
Note that the figure adopts 0-based indexing (e.g., Digit 0 corresponds to the first code position), whereas the formulation in the main text uses 1-based indexing. The embeddings remain broadly distributed, while exhibiting more structured organization after training.}
    \label{fig:code_embeddings}
\end{figure}

We visualize the learned code embeddings on the unit sphere before and after training across different datasets. As shown in Figure~\ref{fig:code_embeddings}, the embeddings remain broadly distributed on the unit sphere after training, while exhibiting more structured patterns. This indicates that training organizes the representation space while preserving its global geometry. These observations qualitatively support the role of semantic grounding.
While SimHash provides an initial coarse-grained semantic structure through token assignments, the alignment objective further refines the embedding space, enabling the model to capture more meaningful semantic relationships.
\setlength{\intextsep}{6pt}
\setlength{\textfloatsep}{6pt}

\subsection{Hyperparameter Analysis on Additional Datasets}
\label{sec:app_hparam}
\begin{figure}[htbp]
    \centering
    \begin{subfigure}{0.32\textwidth}
        \centering
        \includegraphics[width=\linewidth]{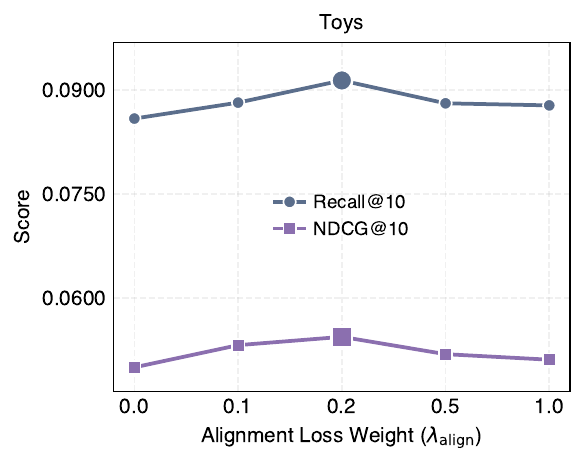}
    \end{subfigure}
    \hfill
    \begin{subfigure}{0.32\textwidth}
        \centering
        \includegraphics[width=\linewidth]{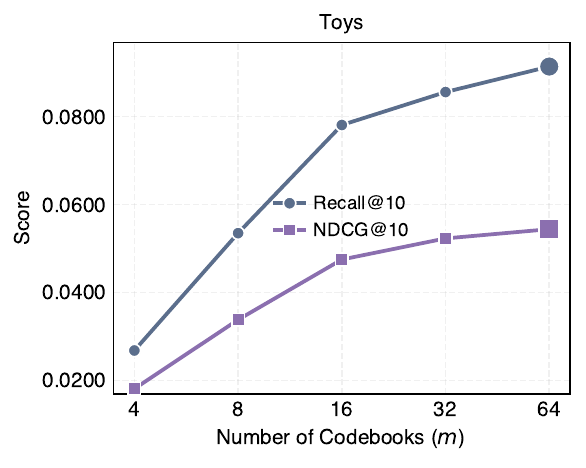}
    \end{subfigure}
    \hfill
    \begin{subfigure}{0.32\textwidth}
        \centering
        \includegraphics[width=\linewidth]{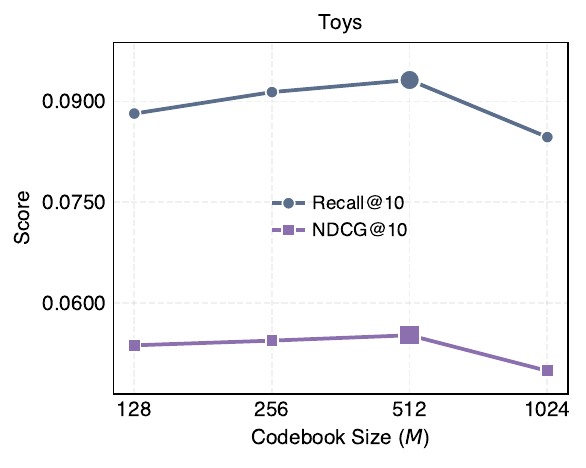}
    \end{subfigure}

    \vspace{0.5em}

    \begin{subfigure}{0.32\textwidth}
        \centering
        \includegraphics[width=\linewidth]{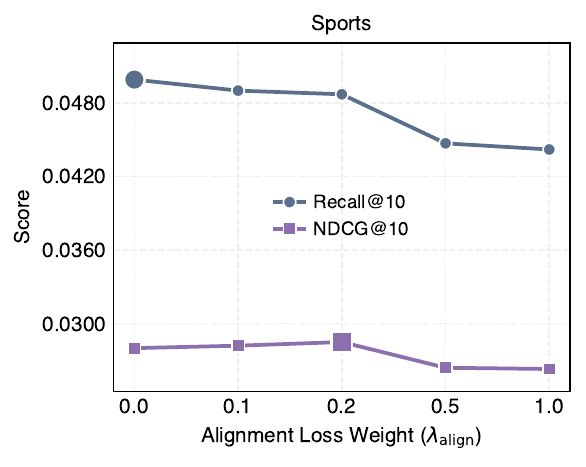}
    \end{subfigure}
    \hfill
    \begin{subfigure}{0.32\textwidth}
        \centering
        \includegraphics[width=\linewidth]{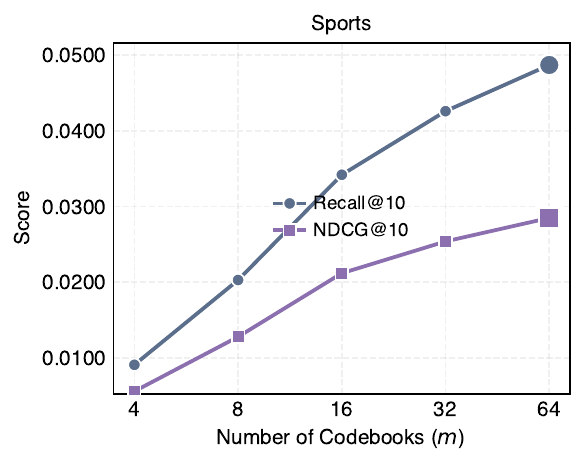}
    \end{subfigure}
    \hfill
    \begin{subfigure}{0.32\textwidth}
        \centering
        \includegraphics[width=\linewidth]{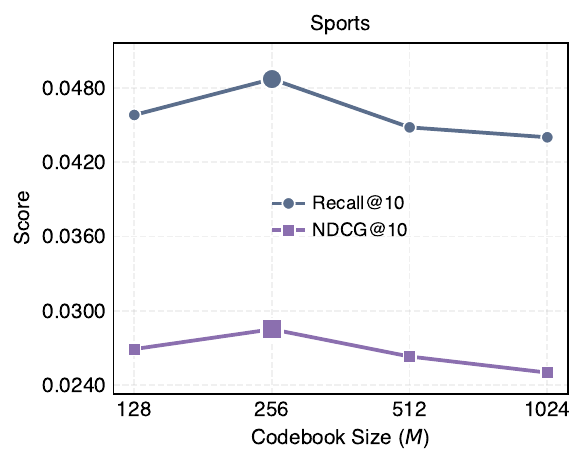}
    \end{subfigure}

    \vspace{0.5em}

    \begin{subfigure}{0.32\textwidth}
        \centering
        \includegraphics[width=\linewidth]{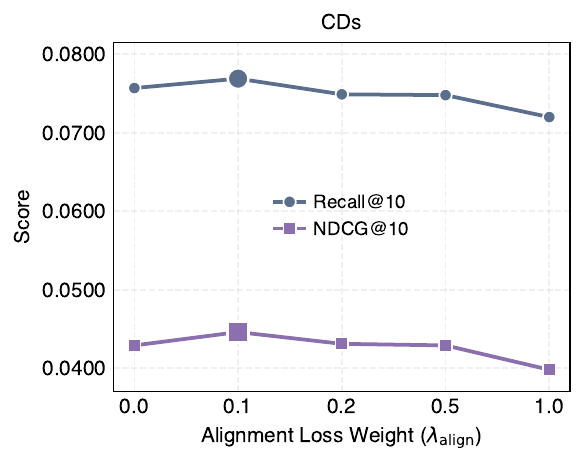}
    \end{subfigure}
    \hfill
    \begin{subfigure}{0.32\textwidth}
        \centering
        \includegraphics[width=\linewidth]{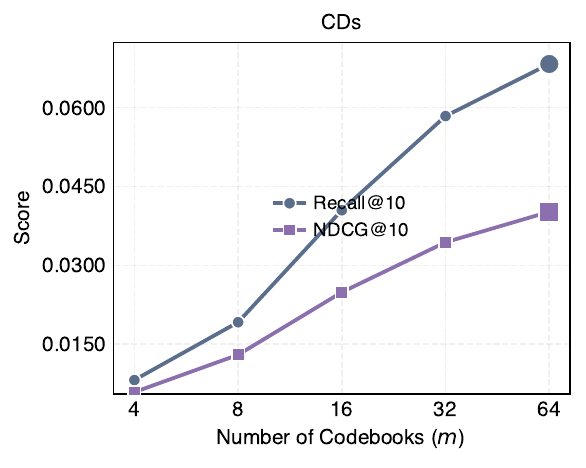}
    \end{subfigure}
    \hfill
    \begin{subfigure}{0.32\textwidth}
        \centering
        \includegraphics[width=\linewidth]{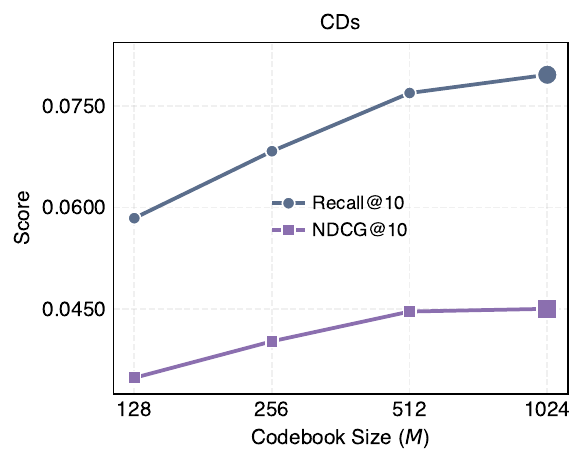}
    \end{subfigure}
    \caption{Performance of \mymodel\ over different hyperparameters on the remaining datasets.}
    \label{fig:app_hparam}
\end{figure}

We report additional hyperparameter analysis results on the remaining datasets in Figure~\ref{fig:app_hparam}, showing that the overall trends are consistent across datasets.
Specifically, moderate values of the alignment weight $\lambda_{\text{align}}$ consistently yield the best performance, whereas overly large values degrade performance. 
Similarly, increasing the number of codebooks improves performance up to a certain point, after which the gains saturate or slightly decline. 
For the codebook size, larger values provide modest improvements initially but exhibit diminishing returns as the size increases.
These results further validate that the effects of key hyperparameters are stable across datasets and support the conclusions drawn in the main text.

\section{Additional Discussions}
\label{sec:additional_discussions}

\subsection{Relation to Diffusion-based Generative Recommendation}
\label{sec:app_diffusion}

Recent diffusion-based generative recommenders explore semantic ID construction for bidirectional and parallel generation.
LLaDA-Rec~\cite{lladarec} identifies a mismatch between residual quantization and bidirectional architectures and replaces hierarchical residual codes with parallel tokenization.
Similarly, DiffGRM~\cite{liu2026diffgrm} introduces Parallel Semantic Encoding to decouple semantic ID positions for bidirectional and parallel generation, while MDGR~\cite{mdgr} adopts a parallel codebook as the structural foundation for masked diffusion.
In contrast, FLASH starts from an existing training-free tokenizer and investigates which decoding paradigm is compatible with its tokenization structure.
Specifically, we revisit the previously reported poor performance of LSH-based tokenization in TIGER~\cite{tiger} and show that pairing its unordered codes with parallel rather than autoregressive decoding substantially improves recommendation performance.

\end{document}